\documentclass[A3paper,11pt]{article}
\pdfoutput=1 

\usepackage{jcappub} 

\usepackage[T1]{fontenc} 
\usepackage{physics}
\usepackage{subcaption}
\usepackage{orcidlink}

\title{\boldmath Cosmological accretion onto braneworld black holes: a relativistic treatment}
\author[a]{Itzi Aldecoa-Tamayo\,\orcidlink{0009-0002-1119-7198},}
\author[a]{Christian~T.~Byrnes\,\orcidlink{0000-0003-2583-6536},}
\author[a]{David Seery\,\orcidlink{0000-0003-3421-6080}}

\affiliation[a]{Astronomy Centre, University of Sussex\\ Falmer, Brighton, BN1 9QH, UK}

\emailAdd{i.aldecoa-tamayo@sussex.ac.uk}
\emailAdd{c.byrnes@sussex.ac.uk}
\emailAdd{d.seery@sussex.ac.uk}

\abstract{Higher-dimensional black holes have been extensively studied over the years, primarily from heuristic and fundamental perspectives or within the context of holographic applications. However, their interaction with ordinary matter confined to the brane is also of particular interest in cosmology. In this work, we revisit accretion within the Randall--Sundrum type II framework, employing the covariant Shiromizu--Maeda--Sasaki formalism together with the Gauss--Codazzi and energy conservation equations. We analyse information propagation in the cosmological fluid and implement a fully relativistic treatment of accretion following Michel's prescription. We find that braneworld effects play a significant role in the early Universe, strongly impacting the evolution of light primordial black holes (PBHs). In particular, the mapping between initial conditions and present-day PBH populations is substantially modified by an extended phase of  early-time accretion that is significantly more efficient than previously found. For certain regions of parameter space, PBHs that could contribute to the present-day dark matter abundance may have formed with masses below the effective four-dimensional Planck scale. The discrepancy between our black hole masses and the most optimistic previous estimates grows as $t^{0.34}$~---a significant difference that reaches up to several orders of magnitude by the end of the strong-accretion epoch, particularly for black holes that form early and for small values of the fundamental Planck scale $M_5$, reaching up to $\sim 10^5$ for the smallest $M_5$ permitted by observations.}

\begin{document}
\maketitle
\flushbottom

\section{Introduction}
\label{sec:intro}
During the late 1990s, a variety of cosmological models incorporating extra spatial dimensions began to emerge, inspired by developments in string theory, attempts to unify the fundamental interactions, and efforts to address the hierarchy problem.
In these theoretical frameworks, gravity extends beyond the familiar four dimensions, propagating through additional spatial directions. These extensions not only alter the behaviour of gravity on small scales, but may also lower the fundamental Planck scale.
Numerous realisations of these higher-dimensional models have been put forward, differing in characteristics such as the number and geometry of the extra dimensions, and in whether the manifold is compact or non-compact.

A particularly influential class of these theories is represented by braneworld models (first introduced in~\cite{Arkani_Hamed_1998} and~\cite{Randall_1999-1,Randall_1999-2}), which describe our observable Universe as a lower-dimensional brane embedded within a higher-dimensional bulk spacetime. In this picture, Standard Model particles are localised on the brane, while gravitational degrees of freedom can propagate into the bulk. This framework leads to significant modifications of both cosmological evolution and the behaviour of black holes on small scales. In this article, we focus on Randall--Sundrum type II (RS-II) geometries~\cite{Randall_1999-2}, in which a single codimension-one brane is embedded in an AdS$_5$ bulk.

Primordial black holes\footnote{For extensive reviews on primordial black holes,
see, for example, refs.~\cite{Green_2021,Escriv__2024,Byrnes:2025tji}.} (PBHs), originally proposed to form in the early Universe~\cite{hawking1971gravitationally}, offer a natural testing ground for such higher-dimensional scenarios. Owing to the variety of their proposed formation channels, PBHs can, in principle, form with masses spanning a wide range. This results in very different properties depending on their horizon radius. In braneworld models, these differences are amplified by the presence of an additional scale: the effective size of the extra dimensions. Black holes with horizon radii significantly smaller than this scale exhibit very peculiar properties, whereas larger black holes are very similar to their $(3+1)$-dimensional counterparts. If black holes originate from formation processes driven by the dynamics of the early Universe, the former are expected to form during a cosmological epoch in which braneworld effects dominate (which we refer to as the ``quadratic regime''), enhancing the differences in their behaviour and evolution relative to the conventional scenario. We refer to them as \emph{small} black holes, and they are the main focus of this work. This paper is part of a series of studies aimed at understanding how a brane-bound observer perceives the influence of a higher-dimensional spacetime on primordial black hole physics. In ref.~\cite{Aldecoa-Tamayo_2026}, we investigated the evaporation of small black holes, and the cosmological and astrophysical signatures of a primordial black hole population in an RS-II universe, whereas here we examine the process of accretion of small black holes.

Primordial black holes in braneworld cosmology have been extensively studied since the early 2000s~\cite{Aldecoa-Tamayo_2026,Guedens_2002-2,Clancy:2003zd,Majumdar_2003,Majumdar:2003sf,Majumdar:2004mz,Majumdar:2005ba,Inoue:2003di,Sendouda:2003dc,Sendouda:2004hz,Tikhomirov:2005bt,Eiroa:2005vd,Keeton:2006di,Sendouda:2006nu,Sendouda:2006yc,Friedlander:2022ttk,Friedlander:2023qmc,ghoshal2025complementaryprobeswarpedextra,vitale2026microscopicprimordialblackholes}. In particular, the mass accretion rate of primordial black holes in a RS-II spacetime geometry was calculated by Guedens, Clancy and Liddle~\cite{Guedens_2002-2}, Majumdar~\cite{Majumdar_2003}, and Tikhomirov and Tsalkou~\cite{Tikhomirov:2005bt}\footnote{The general case of accretion by a $d$-dimensional Schwarzschild black hole was studied in~\cite{John_2013}, where they also consider a $d$-dimensional fluid.}. The first two articles arrive at the same result, except for an undefined parameter $F$ (to which the result is exponentially sensitive) introduced in ref.~\cite{Guedens_2002-2} that measures accretion efficiency, with $F\leq 1$. In both calculations, the authors choose the hydrodynamic approach to describe the accretion process and employ Bondi's approximation~\cite{Bondi1952} for the rate of change of the black hole mass. Their calculations predict a very strong accretion rate in the early stages after formation of small black holes, which clearly diverges from the prediction of standard cosmology where accretion by very light PBHs is expected to be negligible. This surprising result arises from the combination of two factors: a larger accretion radius in comparison to their $(3+1)$-dimensional counterpart and a slower dilution rate with respect to cosmic time for the energy density of the matter fields fluid in early epochs. Later on, Tikhomirov et al. accounted for frequent particle collisions in the cosmological fluid, and argued that this led to a larger value $F\simeq1.54$. The energy density and velocity profiles during the accretion process in braneworlds were also later studied by Harko~\cite{Harko_2009}, who focused on the low-temperature limit and thus do not apply to early epochs.

The central aim of this study is to critically examine the applicability within braneworld models of all standard equations and definitions employed to arrive at the accretion rate $\dot{M}(t)$. To this end, we perform a comprehensive calculation to test the consistency of the adiabatic definition of the information propagation speed in the cosmological fluid. We also complement the discussion in refs.~\cite{Guedens_2002-2,Majumdar_2003,Tikhomirov:2005bt} by recalculating the accretion rate using Michel's relativistic extension of the Bondi accretion formalism~\cite{Michel1972}, \emph{a priori} essential in non-conventional scenarios where modifications to the underlying spacetime geometry may play a significant role. Using additional fundamental conservation equations, we also avoid thermodynamic approximations when evaluating the mass change rate $\dot{M}(t)$ at a particular radius.
 

We also address the validity of Michel's approximation in early epochs, both of which rely on the assumptions of a quasi-steady flow (stationarity) and a negligible self-gravity of the surrounding fluid. This has already been done by Ricotti~\cite{Ricotti_2007} for Bondi's result in the case of PBHs in standard cosmology, but not for braneworlds. It was concluded in ref.~\cite{Ricotti_2007} that it is adequate only for limited ranges of masses in the conventional scenario. In the case of RS-II, we show that it is valid for a much wider parameter range, with the exact domain determined by the collapse dynamics. In our analysis, we  highlight the presence of other forms of gravitational energy surrounding the black hole and their importance in the context of accretion. These components must participate in the accretion process in order to ensure the conservation of energy. In the conventional scenario, most black hole masses require time-dependent self-similar solutions and numerical simulations to accurately describe the accretion flow morphology. In RS-II, these restrictions are softened, allowing for a fully analytic description.

This article is structured as follows. In section~\ref{section-SMS}, we briefly introduce the covariant Shiromizu--Maeda--Sasaki (SMS) framework, which provides the basis for understanding dimensional reduction and the brane-bound four-dimensional perspective. Within this formalism, we discuss the evolution of the cosmological background and examine the implications of such a reduction for small black holes. This is followed by the calculation of the constants of integration in the SMS framework in section~\ref{section-constants-integration}. We then proceed to calculate the propagation scale in the cosmological fluid in section~\ref{section-sound-speed}, required to understand the fluid's flow during the accretion process. In section~\ref{section-accretion}, we derive the energy density and radial four-velocity profiles, and obtain the accretion rate and black hole mass evolution. We also examine the duration of the strong-accretion phase. The validity of Michel’s accretion model in the early Universe is assessed in section~\ref{section-validity}. Finally, section~\ref{section-conclusions} summarises our results.

\vspace{5mm}
\noindent
\textbf{Notation}. We use natural units $c=\hbar=k_B=1$ and the sign convention (--,+,+,+).

\section{The Shiromizu-Maeda-Sasaki framework}\label{section-SMS}
A natural first question in the study of cosmological processes in higher-dimensional spacetimes is whether one should analyse the full higher-dimensional picture or instead perform a covariant reduction to obtain the perspective of an observer confined to the brane. Although the two approaches are formally equivalent, they emphasise different aspects of the physics and are useful in complementary regimes. We choose the latter, since we are particularly interested in the interpretation of the process from the viewpoint of a brane-bound observer. For this purpose, we use the SMS formalism~\cite{Shiromizu_2000}, where the Gauss--Codazzi equations and the Israel--Darmois junction conditions are employed to project the five-dimensional curvature along the brane. We then arrive at the induced field equations on the brane,
\begin{equation}\label{EFE-1}
 ^{(4)}G_{\mu\nu}=\kappa^2T_{\mu\nu}+6\frac{\kappa^2}{\lambda}S_{\mu\nu}-\mathcal{E}_{\mu\nu}
\end{equation}
for the special case of a negligible cosmological constant on the brane $\Lambda_4\simeq 0$ and where
\begin{equation}\label{lambda}
 \lambda=\left(\frac{3}{4\pi}\right)\left(\frac{M_4}{\ell}\right)^2=\left(\frac{3}{4\pi}\right)\left(\frac{M_5^3}{M_4}\right)^2
\end{equation}
is the brane tension, determined by the AdS radius $\ell$ or the fundamental Planck mass $M_5$, $\ell=M_4^2/M_5^3$, with $M_4$ the effective four-dimensional Planck mass. In eq.~\eqref{EFE-1},
$ ^{(4)}G_{\mu\nu}$ is the effective Einstein tensor on the brane, $\kappa$ the four-dimensional Planck scale $\kappa^2=8\pi/M_4^2$,  $T_{\mu\nu}$ the stress-energy tensor of brane matter fields and $S_{\mu\nu}$ and $\mathcal{E}_{\mu\nu}$ braneworld contributions.

The tensor $S_{\mu\nu}$ originates from the extrinsic-curvature contributions to the projected Einstein tensor. In the presence of matter on the brane ($T_{\mu\nu}\neq 0$), the Israel--Darmois junction conditions require $S_{\mu\nu}$  to be non-vanishing in order to ensure geometric consistency. This tensor is fully determined by data on the brane and is quadratic in the stress–energy tensor of the matter fields,
\begin{equation}\label{Smunu-general}
    S_{\mu\nu}=\frac{1}{12}T T_{\mu\nu}-\frac{1}{4}T_{\mu\alpha}T^\alpha_\nu+\frac{1}{24}g_{\mu\nu}\left(3T_{\alpha\beta}T^{\alpha\beta}-T^2\right)\,.
\end{equation}
 Let us consider a perfect fluid of matter fields on the brane. The stress-energy tensor of the matter fields on the brane is of the standard form:
\begin{equation}\label{Tmunu}
    T_{\mu\nu}=(\rho+p)u_\mu u_\nu +p \, g_{\mu\nu}\,,
\end{equation}
with $u_\mu$, $\rho$ and $p$ being the four-velocity, energy density and pressure of the matter content confined to the brane, respectively. For a stress-energy tensor of this kind, eq.~\eqref{Smunu-general} reduces to
\begin{equation}\label{Smunu}
    S_{\mu\nu}=\frac{1}{12}\rho\left[\rho u_\mu u_\nu + (\rho+2p) h_{\mu\nu}\right]\,,
\end{equation}
where $h_{\mu\nu}$ is the spatial part of the brane's metric. In the regime $\rho \gg \lambda$, the tensor $S_{\mu\nu}$ dominates over the matter stress-energy tensor in eq.~\eqref{EFE-1}. Consequently, although $T_{\mu\nu}$ describes the physical matter content on the brane, it is $S_{\mu\nu}$ that primarily governs the spacetime geometry encoded in $^{(4)}G_{\mu\nu}$ in this limit.

The second contribution in the final term in eq.~\eqref{EFE-1} is the projected Weyl tensor $\mathcal{E}_{\mu\nu}$, which encodes contributions from five-dimensional graviton effects. From the perspective of a brane observer, these appear as non-local corrections. A non-vanishing Weyl tensor may be sourced by bulk matter or arise as a backreaction to inhomogeneities on the brane. In particular, gradients in the energy density $\rho$ of the brane matter fields induce a non-zero projected Weyl tensor~\cite{Shiromizu_2000},
\begin{equation}
    \nabla_\mu \rho \neq 0 \quad \to \quad \mathcal{E}_{\mu\nu}\neq 0\,.
\end{equation}
However, the converse need not hold: homogeneity and isotropy of the matter fields do not guarantee $\mathcal{E}_{\mu\nu} = 0$~\cite{Maartens_2004}.

One can interpret the Weyl term as a dark stress-energy tensor and take the fluid Ansatz:
\begin{equation}\label{Emunu}
    \mathcal{E}_{\mu\nu}=-\kappa^2\big(\rho_\mathcal{E}\,u_\mu u_\nu +\frac{1}{3}\rho_\mathcal{E\,}h_{\mu\nu}+\Pi_{\mu\nu}\big)\,,
\end{equation}
where the scalar $\rho_{\mathcal{E}}$ can be interpreted as the energy density of an effective dark fluid residing in the bulk, while $\Pi_{\mu\nu}$ represents an anisotropic stress component.

For example, in the presence of black holes in the bulk, linear perturbations can source Kaluza--Klein modes, which are perceived on the brane as dark radiation, corresponding to $\rho_{\mathcal{E}} \neq 0$. In principle, one may also consider a background with a homogeneous and isotropic (yet non-vanishing) $\rho_{\mathcal{E}}$, while the matter content on the brane remains a perfect fluid with $\nabla_\mu \rho = 0$. However, cosmological constraints require $\rho_{\mathcal{E}}$ to be subdominant at the background level; for instance, $\rho_{\mathcal{E}} / \rho_{\rm rad} \lesssim 0.062$ at the epoch of Big Bang Nucleosynthesis~\cite{Sasankan_2017-1,Sasankan_2017-2}.

We emphasise that the non-local nature of $\mathcal{E}_{\mu\nu}$ implies that the effective gravitational field equations on the brane do not, in general, form a closed system. A complete determination of $\mathcal{E}_{\mu\nu}$ requires solving the bulk gravitational dynamics, particularly in the presence of inhomogeneities. Consequently, assessing its impact in eq.~\eqref{EFE-1} for inhomogeneous and anisotropic configurations is non-trivial.
\\
\\It is customary to write an effective stress energy tensor,
\begin{equation}
 ^{(4)}G_{\mu\nu}=\kappa^2\mathcal{T}_{\mu\nu}\,,
\end{equation}
such that
\begin{equation}\label{Tmunu-eff}
    \mathcal{T}_{\mu\nu}=(\rho_{\rm eff}+p_{\rm eff})\,u_\mu u_\nu + p_{\rm eff}\,g_{\mu\nu}+\Pi_{\mu\nu}\,,
\end{equation}
where the effective quantities are~\cite{Maartens_2004} 
\begin{subequations}\label{effective-quantities}
\begin{equation}\label{rho-eff}  
\rho_{\rm eff}=\rho\left(1+\frac{\rho}{2\lambda}+\frac{\rho_\mathcal{E}}{\rho}\right),
\end{equation}
\begin{equation}\label{p-eff}
    p_{\rm eff}=p+\frac{\rho}{2\lambda}\left(2p+\rho\right)+\frac{\rho_\mathcal{E}}{3}\,.
\end{equation}
\end{subequations}

Depending on the properties of $\Pi_{\mu\nu}$, the effective description may or may not correspond to that of a perfect fluid. We assume that the effective fluid shares the same unit four-velocity $u_\mu$ as in eq.~\eqref{Tmunu}, implying the absence of any relative velocity between the dark radiation and the matter fluid. This assumption is valid for an exact Friedmann–Lemaître–Robertson–Walker (FLRW) brane with an empty bulk~\cite{Maartens_2004}, where symmetry arguments ensure that the dark component is comoving with $u^\mu$.

We emphasise that the effective fluid interpretation can be misleading and should be treated with care. As shown in eq.~\eqref{EFE-1}, in braneworld scenarios---particularly in the high-energy regime $\rho \gg \lambda$—the spacetime curvature sourced by a fluid with energy density $\rho$ and pressure $p$ differs from that in the standard four-dimensional case. One may introduce an effective stress-energy tensor $\mathcal{T}_{\mu\nu}$ which, within conventional $(3+1)$-dimensional general relativity, reproduces the same spacetime geometry; however, this correspondence is purely formal. Consequently, quantities such as the sound speed or the equation of state must be interpreted with caution. It is tempting to introduce an effective equation of state and sound speed based on eqs.~\eqref{effective-quantities}. However, although the modified geometry influences the propagation of pressure-supported disturbances, these perturbations propagate within the physical matter fluid rather than any effective medium. Consequently, an effective sound speed lacks a clear physical meaning: the effective fluid is not a real medium supporting information transfer, but merely a formal construct reproducing the same geometry in standard gravity.

Moreover, the equation of state is intrinsically a thermodynamic quantity, encoding short-range microscopic interactions within the fluid, and should therefore be independent of the spacetime geometry. Introducing effective quantities that do not correspond to physical degrees of freedom can therefore be misleading. While dimensional reduction provides a useful description from the perspective of a brane observer, we refrain from adopting an effective-fluid interpretation in most of this work for these reasons. In section~\ref{section-sound-speed}, we instead show how the additional terms modify the propagation of scalar perturbations and hence the speed at which information propagates within the matter fluid.
%
%
\subsection{Cosmological background in braneworlds}
The cosmological generalisation of RS-II that admits a FLRW foliation results in a first integral, analogous to the first Friedmann equation~\cite{Bin_truy_2000}, of the form
\begin{equation}\label{mod-Friedmann-1}
    H^2=\frac{8\pi}{3M_4^2}\,\rho\left(1+\frac{\rho}{2\lambda}+\frac{\rho_{\mathcal{E}}}{\rho}\right)+\frac{\Lambda_4}{3}-\frac{K}{a^2}\,,
\end{equation}
where $H=\dot{a}/a$ is the Hubble parameter in the brane, with the overdot denoting the cosmic time derivative and $a=a(t)$ the scale factor. $K=0,\pm1$ is the spatial curvature and $\Lambda_4$ the effective cosmological constant on the brane. Equation~\eqref{mod-Friedmann-1} governs the background evolution in the brane.

For early epochs (i.e.~density parameter $\Omega_{\Lambda_4}\ll 1$)
on a flat Friedmann brane ($K=0$), eq.~\eqref{mod-Friedmann-1} can be approximated as
\begin{equation}\label{mod-Friedmann}
    H^2\simeq\frac{8\pi}{3M_4^2}\left(\rho+\frac{\rho^2}{2\lambda}\right)\,,
\end{equation}
where we have set $\rho_{\mathcal{E}}/\rho$ to zero following the observational constraints mentioned above. Equation~\eqref{mod-Friedmann-1} shows that, depending on the magnitude of the energy density, the Universe undergoes different expansion regimes. For the limit $\rho\ll\lambda$, expansion will be dictated by the standard relation $H^2\sim\rho$
(which we describe as the ``linear'' regime), whereas in the limit $\rho\gg\lambda$, it will follow the non-conventional scaling relation $H^2\sim\rho^2$
(which we describe as the ``quadratic'' regime).
The crossover between regimes occurs approximately when $\rho \sim \lambda$,
at a cosmic crossover time $t_c$,
\begin{equation}
    t_c\equiv \frac{\ell}{2}\,.
\end{equation}
Assuming a radiation background with scaling $\rho \propto a^{-4}$,
we have
\begin{subequations}
\begin{equation}\label{quad-reg}
    a\propto t^{1/4}\,,\qquad    \rho=\frac{3M_4^2}{32\pi t_c t}\,,\qquad  H^{-1}=4t
\end{equation}
for early times $t\ll t_c$. Meanwhile,
for $t \gg t_c$ the standard cosmology is recovered,
\begin{equation}\label{linear-reg}
      a\propto t^{1/2}\,,\qquad  \rho=\frac{3M_4^2}{32\pi t^2}\,,\qquad     H^{-1}=2t\,.
\end{equation}
\end{subequations}
Primordial black holes that form deep in the quadratic regime $t\ll t_c$ will have an event horizon $r_0\ll H^{-1}(t_c)\sim \ell$ at formation due to causality arguments. Any dynamical process PBHs undergo during the quadratic regime should still preserve the condition $r_0\ll \ell$.
%
%
\subsection{Black holes for the brane-observer}\label{section-BHs}
Let us first think as five-dimensional observers and consider $\rho=0$ and a small black hole in the brane with curvature radii significantly smaller than the AdS scale. This allows a spherically symmetric geometry and thus a higher-dimensional generalisation of the Schwarzschild metric,\footnote{This solution is known in the literature as the Schwarzschild--Tangherlini metric \cite{Tangherlini:1963bw}.}
\begin{equation}\label{ST-metric-5d}
    ^{(5)}\dd s^2=-\left(1-\frac{r_0^2}{r^2}\right)\dd t^2+\left(1-\frac{r_0^2}{r^2}\right)^{-1}\dd r^2+r^2\dd\Omega_3^2\,,
\end{equation}
with $\Omega_n$ the area of a unit $n$-sphere and $r_0$ the black hole event horizon radius,
\begin{equation}\label{horizonradius}
    r_0=\sqrt{\frac{8}{3\pi}}\frac{\sqrt{\ell M}}{M_4}\,,
\end{equation}
where $M$ is the global mass of the black hole. When we examine the Ricci tensor $^{(5)}R_{ab}$ of the full five-dimensional geometry,\footnote{Greek indices will be used for quantities on the brane, and Latin indices for those in the bulk.} we see that \eqref{ST-metric-5d} is a solution to the vacuum Einstein equations
\begin{equation}
    ^{(5)}G_{ab}=0\,.
\end{equation}
Now consider what is seen by a brane-bound observer. The projection of the metric~\eqref{ST-metric-5d} on the brane in Gaussian normal coordinates is
\begin{equation}\label{ST-metric-4d}
       ^{(4)}\dd s^2=-\left(1-\frac{r_0^2}{r^2}\right)\dd t^2+\left(1-\frac{r_0^2}{r^2}\right)^{-1}\dd r^2+r^2\dd\Omega_2^2\,.
\end{equation}
If we now calculate the effective Ricci tensor $^{(4)}R_{\mu\nu}$ of the brane, we see that it does not vanish for~\eqref{ST-metric-4d}. The black hole metric from the perspective of a brane-bound observer is a solution to the trace-free Einstein equation, but
\begin{equation}\label{nonvanish-Rmunu}
    ^{(4)}G_{\mu\nu}=\,^{(4)}R_{\mu\nu}\neq 0\,.
\end{equation}
It is customary, when dealing with solutions of this type, to interpret $^{(4)}R_{\mu\nu}$ as a stress-energy tensor. In the present case, however, there is no matter content on the brane; $^{(4)}R_{\mu\nu} \neq 0$ arises purely from dimensional reduction and is therefore geometrical in origin. We thus identify $^{(4)}R_{\mu\nu} = \mathcal{E}_{\mu\nu}$ and adopt the fluid Ansatz given in eq.~\eqref{Emunu}.

To characterise the morphology of the gravitational “fluid” around the black hole as seen by a brane observer, we impose the traceless condition for a Schwarzschild--Tangherlini spacetime, yielding
\begin{equation}\label{rhoE-asr0}
    \rho_\mathcal{E}^{\small\rm BH}=-\frac{1}{\kappa^2}\frac{r_0^2}{r^4}\,,
\end{equation}
or, expressed in terms of the mass,
\begin{equation}\label{rhoE-vacuum}
    \rho_\mathcal{E}^{\small\rm BH}=-\frac{\ell M}{3\pi^2r^4}\,,
\end{equation}
and 
\begin{equation}\label{Pimunu}
    \Pi_{\mu\nu}=\text{diag}\bigg[0,-\frac{1}{3}\rho_\mathcal{E}^{\small\rm BH}g_{rr}+\bigg(\frac{r_0}{r}g^{rr}\bigg)^2,-\frac{1}{3}\rho_\mathcal{E}^{\small\rm BH}g_{\theta\theta},-\frac{1}{3}\rho_\mathcal{E}^{\small\rm BH}g_{\phi\phi}\bigg]\,,
\end{equation}
where we have taken $\Pi_{\mu\nu}$ to be diagonal due to the symmetries of the underlying metric. 
We find that, if $\Pi_{\mu\nu}$ is written in the form
\begin{equation}
\Pi_{\mu\nu}=\pi_{\mathcal{E}}\bigg(\frac{1}{3}h_{\mu\nu}-r_\mu r_\nu\bigg)
\end{equation}
for some function $\pi_{\mathcal{E}}(r)$ and where $r_\mu$ denotes the unit radial vector, one obtains the relation
\begin{equation}\label{eos-weyl}
    \pi_{\mathcal{E}}=-\rho_\mathcal{E}^{\small\rm BH}\,.
\end{equation} 
This can be interpreted as an equation of state for the Weyl term with $w_\mathcal{E}=-1$, as done by Gregory et al.~\cite{Gregory_2004}. In general, $\rho_{\mathcal{E}}^{\small\rm BH}$ need not satisfy this relation in realistic scenarios. However, for an isolated black hole in an empty five-dimensional spacetime, eq.~\eqref{eos-weyl} is expected to hold, while alternative equations of state may characterise other simplified configurations.

We can now re-express the metric~\eqref{ST-metric-4d} in terms of the dark radiation energy density,
\begin{equation}\label{metric-cloud}
   \dd s^2=-\left(1+\frac{8\pi}{M_4^2}\rho_\mathcal{E}^{\small\rm BH}\,r^2\right)\dd t^2+\left(1+\frac{8\pi}{M_4^2}\rho_\mathcal{E}^{\small\rm BH}\,r^2\right)^{-1}\dd r^2+r^2\dd \Omega^2_2\,,
\end{equation}
where we emphasise that $|\rho_\mathcal{E}^{\small\rm BH}|$ decreases radially, and that $\rho_\mathcal{E}^{\small\rm BH} < 0$, as required for an attractive gravitational potential.

Equation~\eqref{metric-cloud} clarifies the interpretation of the Schwarzschild--Tangherlini solution for a brane-bound observer. In this picture, one observes a spherically symmetric cloud of dark radiation, with energy density profile $\rho_\mathcal{E}^{\small\rm BH}(r)$ determined by the five-dimensional parameter $M$ and given by eq.~\eqref{rhoE-vacuum}. The profile decreases rapidly with radius, featuring an event horizon at $r = r_0$ and a singularity at $r = 0$.
\section{Conservation equations and constants of integration}\label{section-constants-integration}
We now turn to the conservation equations governing the dynamics of systems in the Randall--Sundrum type II framework. 
In the absence of energy-momentum exchange between the bulk and the brane, the standard four-dimensional conservation equations remain valid,
\begin{equation}\label{cons-eq-1}
    \nabla^\nu T_{\mu\nu}=0\,,
\end{equation}
implying that the stress–energy tensor of the matter fields on the brane is conserved independently. Meanwhile, the four-dimensional contracted Bianchi identities require
\begin{equation}
    \nabla^\nu \,^{(4)}G_{\mu\nu}=0\,,
\end{equation} 
which, upon application to eq.~\eqref{EFE-1}, give rise to an additional set of conservation equations,
\begin{equation}\label{NLCE}
    \nabla^\nu \mathcal{E}_{\mu\nu}=\frac{6\,\kappa^2}{\lambda}\nabla^\nu S_{\mu\nu}\,.
\end{equation}
Note that from equations~\eqref{Smunu} and \eqref{Emunu}, we can see that eq.~\eqref{NLCE} encodes how spacetime variations of the brane matter content act as a source for dark radiation. 
This means that we must ensure consistency: ordinary matter must evolve so that energy-momentum is conserved, but it must do so in a way that does not violate geometric constraints imposed by~\eqref{NLCE}. The Bianchi identities ensure consistency in the evolution of the geometry of the Universe. 

Note that eq.~\eqref{NLCE} implies a mutual coupling: non-local bulk dynamics can also affect the distribution and evolution of the matter fluid on the brane, and this effect is encoded solely in eq.~\eqref{NLCE}.

There is an additional conservation equation that is essential for a complete description of the accretion flow. It enforces conservation of energy along the fluid flow and is obtained by projecting the energy–momentum conservation law along the fluid four-velocity,
\begin{equation}\label{proj-cons}
    u_\mu\, \nabla_\nu T^{\mu\nu}=0\,.
\end{equation}

In the remainder of this section, we identify useful constants of integration arising from the above conservation equations. Since our analysis is formulated within the framework of Michel accretion---which assumes a quasi-steady flow and, consequently, a time-independent spacetime geometry---we adopt the following general static metric on the brane
\begin{equation}\label{general-metric}
    \dd s^2=-f(r)\dd t^2+\frac{1}{f(r)}\dd r^2+r^2\dd\Omega_2
\end{equation}
and fix our coordinate system so that the fluid four-velocity takes the form
\begin{equation}\label{4-velocity}
    u^{\mu}=(u^t,u,0,0)\,,
\end{equation}
where, in the context of accretion, we expect $u<0$, corresponding to inward flow towards the black hole, while we take $u^t>0$. Imposing the normalisation condition $u^\mu u_\mu=-1$, consistent with the sign convention adopted in this work, for the metric~\eqref{general-metric} the four-velocity vector takes the form
\begin{equation}\label{coordinates}
    u^\mu=\bigg(\frac{\sqrt{f(r)+u^2}}{f(r)},u,0,0\bigg)\,.
\end{equation} 
We distinguish between constants of integration associated solely with brane quantities, which we refer to as \emph{local} constants of integration, and those that depend on bulk degrees of freedom, which we term \emph{non-local} constants of integration.
\subsection{Local constants of integration}
To extract the first constant of integration, we evaluate the time component of the conservation equation \eqref{cons-eq-1},
\begin{equation}\label{partial-A1}
    \partial_r\left(\sqrt{-g}T^r_t\right)=0\,,
\end{equation}
where $g$ is the determinant of the metric, i.e.~$\sqrt{-g}=r^2\sin\theta$. We  define the first constant $A_1$ by integrating~\eqref{partial-A1} over all angles,
\begin{equation}\label{pre-A1}
    A_1=\oint \sqrt{-g}T^r_t\;\text{d}\theta\,\text{d}\phi=4\pi r^2\,T^r_t\,,
\end{equation}
such that $\partial_r A_1=0$. For a perfect fluid, eq.~\eqref{pre-A1} takes the form
\begin{equation}\label{A1}
A_1=4\pi\big(\rho(r)+p(r)\big)\,u(r)\,r^2\sqrt{f(r)+u^2(r)}\,.
\end{equation}
The second constant of integration is obtained from~\eqref{proj-cons},
\begin{equation}
    \partial_r\left(\ln(ur^2)+\int^\rho\frac{\dd\tilde{\rho}}{\tilde{\rho}+p(\tilde{\rho})}\right)=0\,.
\end{equation}
We make the choice $\rho(r\to\infty)=\rho_\infty$ for the lower boundary in the integral for convenience and define
\begin{equation}\label{A2}
    A_2=ur^2\exp\left(\int_{\rho_\infty}^\rho\frac{\dd\tilde{\rho}}{\tilde{\rho}+p(\tilde{\rho})}\right)\,.
\end{equation}
These constants of integration are the main ingredients to study the velocity and energy density profiles surrounding the black hole during the accretion process, crucial to determine the mass change rate of the black hole.  
%
%
\subsection{Non-local constants of integration}
We now focus on the conservation equation~\eqref{NLCE}. Using eqs.~\eqref{Smunu} and~\eqref{Emunu}, we can write~\eqref{NLCE} explicitly,
\begin{equation}\label{NLC-1}
    \nabla_\mu\bigg[\frac{\kappa^2}{2\lambda}\rho\big(\rho u^\mu u_\nu + (\rho+2p) h^\mu_{\nu}\big)+\kappa^2\big(\rho_\mathcal{E}u^\mu u_\nu +\frac{1}{3}\rho_\mathcal{E}h^\mu_\nu+\Pi^\mu_{\nu}\big)\bigg]=0\,.
\end{equation}
Adopting again a stationary configuration in which the energy densities, metric, and velocity fields depend only on $r$, and using $h^r_t = u^r u_t$, the time component ($\nu = t$) of eq.~\eqref{NLC-1}, integrated over all angles yields
\begin{equation}\label{Cons-eq-2-b}
   \partial_r\bigg[4\pi r^2\bigg(\frac{\rho(\rho+p)}{\lambda}+\frac{4}{3}\rho_\mathcal{E}\bigg) u u_t\bigg]=0\,,
\end{equation}
where we have taken $\Pi^\mu_\nu$ to be diagonal due to the symmetries of the system and $\Pi_{tt}=0$, just as in section~\ref{section-BHs}. Dividing eq.~\eqref{Cons-eq-2-b} by $A_1$, results in a new constant of integration,
\begin{equation}\label{A3}
   A_3=\frac{\rho}{\lambda}+\frac{4}{3}\frac{\rho_\mathcal{E}}{\rho+p}\,.
\end{equation}
\section{Information propagation speed}\label{section-sound-speed}
In compressible fluid dynamics, a distinguished role is played by the information propagation speed associated with pressure-supported disturbances. Since pressure gradients provide the only mechanism by which spatially separated fluid elements can influence one another, the propagation of small-amplitude compressive perturbations determines the causal structure of the flow. In accretion problems this has a direct physical interpretation: comparing the bulk flow velocity to the information propagation speed determines whether perturbations can propagate upstream or are irreversibly advected downstream, a distinction that underlies the definition of the sonic point and the global structure of Bondi--Michel solutions, which marks the crossover.

A general characterisation of information transport is obtained by analysing linear perturbations of the flow. For a given background configuration, perturbations obey a dispersion relation $\omega(k)$ and the propagation of localised disturbances is governed by the group velocity $v_g=\dd \omega/\dd k$. In the conventional scenario, this group velocity coincides with the characteristic speeds of the first-order Euler equations and with the square-root of the thermodynamic derivative $(\dd p/\dd \rho)_s$ evaluated at constant entropy $s$, known as the sound speed, $c_s$. However, in more general situations---such as dispersive effective theories, higher-derivative models, or fluids with additional dynamical degrees of freedom---the perturbation equations need not be first order, the dispersion relation may be non-linear in $k$, and the information propagation speed is no longer determined by the thermodynamic quantity $c_s$. In such cases, the relevant propagation scale must be computed directly from the perturbation spectrum, with the group velocity (and, where appropriate, its high-frequency limit) providing the appropriate measure of information transport.\footnote{See Brillouin's discussion (ref.~\cite{Brillouin1960}) on information/perturbation propagation speeds.} 

To investigate the propagation of information in a fluid in braneworld scenarios, we study the dynamics of scalar perturbations around the background solution by means of the perturbed Einstein equations. A crucial aspect of perturbation theory in braneworld models is that consistency requires perturbing both the metric and the stress-energy tensor simultaneously. In conventional cosmology, perturbing only the matter sector over a fixed background is often a valid approximation, since the coupling between matter and metric perturbations is weak enough to be neglected at leading order. However, in braneworld scenarios the intrinsic coupling between matter fluctuations on the brane and five-dimensional gravitational perturbations in the bulk geometry demands greater care: neglecting either sector leads to an incomplete description, as the perturbed Einstein equations enforce their mutual backreaction. In fact, restricting the analysis to matter perturbations alone artificially suppresses this coupling and recovers the standard result $v_g=c_s$, as if the extra dimension were absent. Carrying out the full perturbation analysis, as we do in this section, yields instead a third-order wave equation that encodes the brane–bulk coupling. Nevertheless, in the quadratic-regime and sub-horizon limits, the solutions of this richer equation reduce to the conventional result, confirming that the effect of the bulk is suppressed at small scales

Scalar perturbations in braneworld scenarios have been studied extensively in the literature (see the references collected in ref.~\cite{Maartens_2010}). One of the main complications is that, unlike at the background level---where the projected Weyl tensor can be neglected and the dynamics can be understood locally---there is no physical motivation to set its perturbations to zero. As a result, scalar perturbations on the brane generically couple to perturbations in the bulk, leading to a system of equations that is not closed. In certain regimes, however, suitable approximations allow the system to be closed, yielding a complete description governed solely by brane quantities (e.g.~\cite{Langlois_2001}).

A complete study of scalar perturbations in braneworld cosmology was performed in ref.~\cite{Cardoso_2007}, and we follow their approach to derive the corresponding wave equation. They make use of a gauge invariant master variable $\Omega$ previously introduced in refs.~\cite{Mukohyama_2000,Kodama_2000}, where it was shown that scalar-type perturbations of the bulk geometry are governed solely by this variable. Brane perturbations can be expressed in terms of gauge-invariant quantities, as in the conventional scenario, with the most relevant for this work being the density contrast $\Delta$ and the entropy perturbation $\Gamma$,
\begin{equation}
    \rho\Delta=\delta\rho-3H\delta q,\qquad \Gamma=\delta p-c_s^2\delta\rho\,,
\end{equation}
where $\delta q$ is the velocity perturbation.

The connection between the bulk master variable and the brane quantities is obtained by perturbing the junction conditions at the brane. This imposes the following condition for $\Omega$ when evaluated on the brane:
\begin{equation}\label{boundary-cond}
    \left[\partial_n\Omega+\frac{1}{\ell}\left(1+\frac{\rho}{\lambda}\right)\Omega+\frac{6\rho a^3}{\lambda k^2}\Delta\right]_b=0\,,
\end{equation}
with $\partial_n$ being the directional derivative orthogonal to the brane and where a harmonic basis for the perturbations is used, 
\begin{equation}
    \delta T_0^0=-\delta\rho \exp(i\mathbf{k\cdot x}),\qquad \delta \mathcal{E}^0_0=\kappa^2\delta\rho_{\mathcal{E}}\exp(i\mathbf{k\cdot x})\,.
\end{equation}
Perturbing the Einstein field equations on the brane, the effective stress-energy tensor conservation equations and the junction conditions on the brane (see ref.~\cite{Cardoso_2007} for more details), result in a wave equation for the matter density contrast  
\begin{subequations}\label{wave-eq}
    \begin{equation}
\Delta''+\left(1-3w\right)aH\Delta'+\left(k^2c_s^2+\frac{3\rho a^2}{\ell^2\lambda}A+\frac{3\rho^2a^2}{\ell^2\lambda^2}B\right)\Delta=-\frac{k^2\Gamma}{\rho}+\left(\frac{k^4(1+w)}{3\ell a^3}\right)\Omega_b
 \end{equation}
 \begin{equation}
    A=3w^2-2w-1\qquad B=-6w-4\,,
 \end{equation}
\end{subequations} 
where $w$ is the ratio 
 \begin{equation}\label{baro-w}
     w=\frac{p}{\rho}\,.
 \end{equation}
Our result matches the solution obtained in ref.~\cite{Cardoso_2007} for the particular case $w=c_s^2$.
\subsection{Barotropic radiation fluid}
Both in the background and in the classical Michel accretion framework, the flow is expected to be adiabatic with entropy conserved along streamlines. The thermodynamic closure may therefore be taken to be $p=p(\rho)$. We further make the assumption of a barotropic equation of state (i.e.~$w$ in~\eqref{baro-w} is a constant) during the quadratic regime. 
Combining the energy–momentum conservation of ordinary matter
\begin{equation}
    \dot{\rho}=-3(1+w)H\rho
\end{equation}
with the modified Friedmann equation~\eqref{mod-Friedmann}, allows us to find the scaling of the scale factor and Hubble parameter with conformal time in the $\rho/\lambda\gg1$ approximation,
\begin{equation}
    a\propto \eta^{1/(2+3w)},\qquad H=\frac{1}{(2+3w)a\eta}\,.
\end{equation}
In this regime, the directional derivative $\partial_n$ in~\eqref{boundary-cond} happens to depend solely on brane quantities (see equation 4  in ref.~\cite{Cardoso_2007}), and eq.~\eqref{boundary-cond} becomes
\begin{equation}\label{boundary-cond-1}
    \frac{\dd\Omega_b}{\dd\eta}\simeq\frac{1}{3\eta}\left(1+\frac{3a\eta}{\ell}\right)\Omega_b+\frac{2\ell a^3}{k^2\eta}\Delta\,.
\end{equation}
This results in eqs.~\eqref{wave-eq} and~\eqref{boundary-cond-1} forming a closed system of equations on the brane, which can be combined into a single third-order differential equation of $\Delta(\eta)$, independent of the quantity $\Omega_b$.

We then plug in the plane wave Ansatz: 
\begin{equation}
    \Delta=e^{i(kx-\omega\eta)}
\end{equation}
and solve for $\omega$. The group velocity is then defined by
\begin{equation}
    v_g=\frac{\dd \omega}{\dd k}\,.
\end{equation}
We obtain three branches: the first two are equal but opposite in sign, $\pm\omega^{(1)}(k)$, and with all imaginary terms being suppressed on sub-horizon scales, $(k\eta)^{-1}\to 0$; the third solution $\omega^{(3)}(k)$ is purely imaginary (i.e.~not oscillatory) and fully suppressed in sub-horizon scales. Since the imaginary part of $\omega^{(1)}(k)$ is suppressed
\begin{equation}
    \frac{\Im(\omega^{(1)}(k))}{\Re(\omega^{(1)}(k))}\ll1\,,
\end{equation}
the solution is a well-defined wavepacket. 

We compute the group velocity by differentiating the real part of $\omega^{(1)}(k)$ with respect to $k$ and obtain
\begin{equation}
    v_g=\frac{\dd \omega^{(1)}}{\dd k}\simeq \sqrt{w}+\left[f_1(w)+f_2(w,\eta)\left(\frac{a\eta}{\ell}\right)\right]\left(\frac{1}{k\eta}\right)^2
\end{equation}
up to second order in the expansion around $(k\eta)^{-1}\to0$, with
\begin{equation}
f_1=-\frac{315w^4-318w^3-313w^2+152w+48}{72w^{5/2}\,(2+3w)^2},
\end{equation}
\begin{equation}
    f_2=-\frac{27w^3-12w^2+w+4}{6w^{3/2}\,(2+3w)}\,\eta^{\frac{1}{2+3w}-\frac{1}{3}}.
\end{equation}
We highlight that the dimensionless quantity $(a\eta/\ell)$ is a suppression factor for any $\eta$ during the quadratic regime, imposed by the condition $t/t_c\ll 1$ in cosmic time. Due to both suppression factors, we see that 
\begin{equation}
    v_g\simeq \sqrt{w}\,.
\end{equation}
In the adiabatic scenario during the quadratic regime,
for an isotropic plasma dominated by ultra-relativistic degrees of freedom,
the appropriate choice of equation of state is $w=1/3$, given an isotropic plasma. In that case, the group velocity is
\begin{equation}
    v_g\simeq \frac{1}{\sqrt{3}}-\left[\frac{7}{3\sqrt{3}}+\frac{2}{\sqrt{3}}\left(\frac{a\eta}{\ell}\right)\right]\left(\frac{1}{k\eta}\right)^2.
\end{equation}
We conclude that for a barotropic fluid in the quadratic regime, the approximation
\begin{equation}
    v_g\simeq c_s
\end{equation}
is reasonable in sub-horizon scales.

In previous literature \cite{Maartens_2004}, an effective sound speed, defined as $c_{s,\rm eff}^2=\partial p_{\rm eff}/\partial \rho_{\rm eff}$, was introduced. In the quadratic regime, this quantity evaluates to $c_{s,\rm eff}^2 = 5/3$ (corresponding to $w = 1/3$), which may naively suggest a superluminal propagation and thus a potential violation of causality. However, this interpretation is misleading: the quantity does not correspond to a physical signal propagation speed. We conclude that perturbations within the fluid propagate subluminally in the quadratic regime, and causality remains preserved. Even though the underlying geometry differs from the conventional scenario, the corrections to the group velocity are negligible in sub-horizon scales.
%
\section{Michel accretion}\label{section-accretion}
We adopt Michel's relativistic formalism to model accretion within our framework. This formulation relies on several assumptions that restrict its validity to particular regions of parameter space. We discuss the validity of these approximations in section~\ref{section-validity}.

Michel accretion follows directly from energy conservation---specifically, eq.~\eqref{cons-eq-1}---and the constant of integration $A_1$ acquires the physical interpretation of the black hole mass accretion rate, $A_1\equiv-\dot{M}$. Under the assumptions taken in this accretion model, $\dot{M}$ must remain constant for all values of radial coordinate $r$. 

 Beyond this standard identification, the braneworld framework introduces an additional conserved quantity, given by eq.~\eqref{Cons-eq-2-b},
\begin{equation}\label{Cons-eq-2-c}
    \partial_r\left(A_3\,\dot{M}\right)=0\,.
\end{equation}
Here, $A_3$, defined in eq.~\eqref{A3}, plays a role analogous to that of $A_1$: just as $A_1\equiv-\dot{M}$
 encodes the transformation of the energy-momentum tensor $T_{\mu\nu}$ of the accreted fluid into the black hole mass growth rate, $A_3$ encodes the transformation of the gravitational energy stored in $S_{\mu\nu}$ and the component of $\mathcal{E}_{\mu\nu}$ sourced by gradients in the energy density profile, into the growing Weyl cloud of the Schwarzschild--Tangherlini black hole present in the covariant four-dimensional SMS description. Crucially, $A_3$ is conserved alongside the mass accretion rate, and in the late-time limit it drives the Weyl tensor $\mathcal{E}_{\mu\nu}$
toward the form~\eqref{Emunu}, with $\rho_{\mathcal{E}}$ and $\Pi_{\mu\nu}$ as given in eqs.~\eqref{rhoE-asr0} and~\eqref{Pimunu}, respectively.

This is the distinctive feature of accretion in braneworlds: a conserved quantity that concurrently transforms the gravitational energy content of the brane. Since $S_{\mu\nu}$ and $\mathcal{E}_{\mu\nu}$ are not fundamental degrees of freedom but rather artifacts of the projection of the 5D bulk geometry onto the brane, they cannot drive the growth of the black hole mass, but they must transform consistently, with the Weyl cloud---itself a purely four-dimensional construct as well---evolving according to
\begin{equation}
\dot{\rho}_\mathcal{E}^{\small\rm BH}=-\frac{\ell \dot{M}}{3\pi^2r^4}\,,
\end{equation}
as given by eq.~\eqref{rhoE-vacuum}. This is required by energy conservation: were these projection artifacts to evolve without constraint, the Weyl cloud would grow in lockstep with $\dot{M}$ without any physical justification. Since $M$ and $\rho$ are physical observables that exist independently of the dimensional reduction and must remain consistent between the bulk and brane descriptions by construction of the SMS framework, the extra terms in the Einstein field equations must transform accordingly---and it is precisely this transformation that eq.~\eqref{Cons-eq-2-c} governs.

In the following, we close the discussion of the braneworld-specific conservation structure and turn our attention to the physically observable accretion rate $\dot{M}$, which is the quantity of primary interest from a braneworld cosmology standpoint. To that end, we evaluate the energy density and four-velocity profiles in the vicinity of the black hole in order to determine $\dot{M}$ at a scale $r$ where all relevant quantities are known.
\subsection{Critical radius in Schwarzschild-Tangherlini spacetimes}
Let us combine the radial component of the conservation equation~\eqref{cons-eq-1},
\begin{equation}\label{radial-cons-1}
    uu'+\frac{f'(r)}{2}=-\frac{p'}{\rho+p}\left(f+u^2\right)\,,
\end{equation}
with eq.~\eqref{proj-cons}, \footnote{At this stage, it is customary to use rest-mass current conservation, $\nabla_\mu (\rho_0 u^\mu) = 0$ (e.g.\ ref.~\cite{Tikhomirov:2005bt}), with $\rho_0$ the rest-mass density, instead of the projection along the four-velocity. However, this cannot be used in the quadratic regime because the fluid is expected to be highly relativistic.}
\begin{equation}\label{cons-der}
    \frac{\rho'}{\rho+p}+\frac{u'}{u}+\frac{2}{r}=0
\end{equation}
in order to isolate $u'/u$ and $\rho'/\rho$,
\begin{equation}
    \frac{\rho'}{\rho}=\frac{\mathcal{D}_1}{\mathcal{D}_2}\,,\qquad \frac{u'}{u}=\frac{\mathcal{D}_3}{\mathcal{D}_4}\,,
\end{equation}
with
\begin{subequations}
   \begin{equation}
    \mathcal{D}_1=\frac{2u^2}{r}-\frac{f'(r)}{2}\,,
\end{equation}
\begin{equation}\label{eq-D-2}
    \mathcal{D}_2=\frac{wf+(w-1)u^2}{w+1}\,,
\end{equation}
\begin{equation}
    \mathcal{D}_3=\frac{2w(f+u^2)}{r}-\frac{f'(r)}{2}\,,
\end{equation}
\begin{equation}\label{eq-D-4}
    \mathcal{D}_4=-\left(wf+(w-1)u^2\right)\,.
\end{equation}
\end{subequations}
Let us study eq.~\eqref{eq-D-2} at the boundaries. We evaluate $\mathcal{D}_2$ at the horizon $r=r_0$,
\begin{equation}\label{lim-r0}
    \mathcal{D}_2(r_0)=\frac{w-1}{w+1}\,u^2\,,
\end{equation}
and in the asymptotic region,
\begin{equation}\label{lim-inf}
    \mathcal{D}_2(r\to\infty)=\frac{w}{w+1}\,.
\end{equation}
For $0< w\leq 1$, eq.~\eqref{lim-r0} is non-positive and~\eqref{lim-inf} is non-negative. Therefore, $\mathcal{D}_2$ must vanish at some critical radius $r_{\rm cr}$. For the flow to be regular, we must impose $\mathcal{D}_1(r_{\rm cr})=\mathcal{D}_2(r_{\rm cr})=0$  (see ref.~\cite{shapiro-book} for a pedagogical discussion). The same applies for the other two equations, i.e.~$\mathcal{D}_3(r_{\rm cr})=\mathcal{D}_4(r_{\rm cr})=0$. Both conditions provide the equality
  \begin{equation}\label{eqs-csuc}
   u_{\rm cr}^2=\frac{wf_{\rm cr}}{1-w}=
\frac{f_{\rm cr}'r_{\rm cr}}{4}\,,
\end{equation}
that allows us to define quantities at the critical radius. For the Schwarzschild--Tangherlini metric,
    \begin{equation}\label{eqs-as-w}
r_{\rm cr}^2=\frac{(1+w)\,r_0^2}{2w}\,,
\qquad
    u_{\rm cr}^2=\frac{w}{1+w}\,.
\end{equation}
Based on our conclusions from section~\ref{section-sound-speed}, eq.~\eqref{eqs-as-w} implies that the condition
\begin{equation}\label{subsonic-rc}
u_{\rm cr}^2<c^2_{s,c}
\end{equation}
is met. Unlike the conventional adiabatic scenario, where $u_{\rm cr}^2=c_{s,c}^2$, in Schwarzschild--Tangherlini spacetimes the fluid is subsonic at the critical point. Whereas the critical point has no particular physical interpretation beyond an algebraic requirement for continuity, the sonic point $r_s$ where $u_{\rm cr}^2=c_{s,c}^2$, dictates where the maximum propagation speed of perturbations becomes comparable to the bulk velocity of the infalling flow. In the supersonic region, perturbations cannot propagate outward (upstream), whereas in the subsonic region, information can travel in both directions. In this geometry, $r_{\rm cr}>r_s>r_0$. 
\begin{figure}
    \centering
    \includegraphics[width=0.5\linewidth]{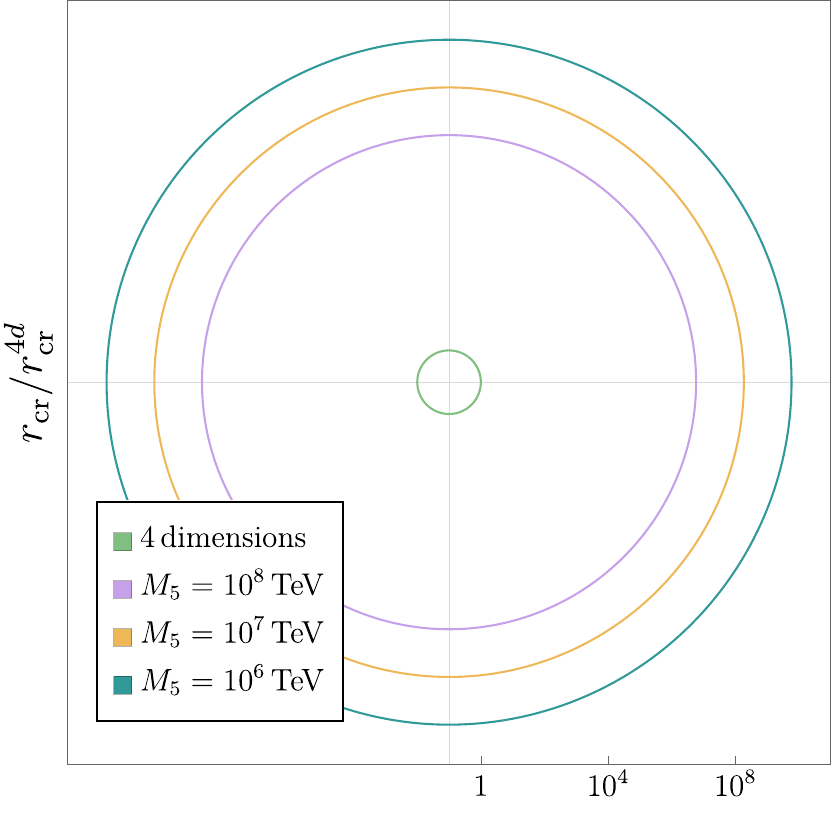}
    \caption{Accretion scale for a fixed mass ($M=10^{10}M_4$) for different fundamental scales $M_5$ (with $M_5=10^{6}\,\text{TeV}$ being the smallest value allowed by observations~\cite{Kapner:2006si}), compared to the $(3+1)$-dimensional Schwarzschild case where the critical radius takes the value $r_{\rm cr}^{4d}=3M$ \cite{Michel1972}. In both cases, we assume an equation-of-state parameter $w = 1/3$.}
    \label{fig:accretion-scale}
\end{figure}
The critical radius is essential for determining the accretion rate $\dot{M}$ and the overall structure of the inflow. Evaluating the constants of integration~\eqref{A1} and~\eqref{A2} at $r_{\rm cr}$ allows us to define the energy density at the critical point,
\begin{equation}\label{rho-c}
    \rho_{\rm cr}=\left(1+w\right)^{\frac{1+w}{2w}}\rho_\infty\,.
\end{equation}
\subsection{Velocity and energy density profiles}
\begin{figure}
    \centering
    \includegraphics[width=1\linewidth]{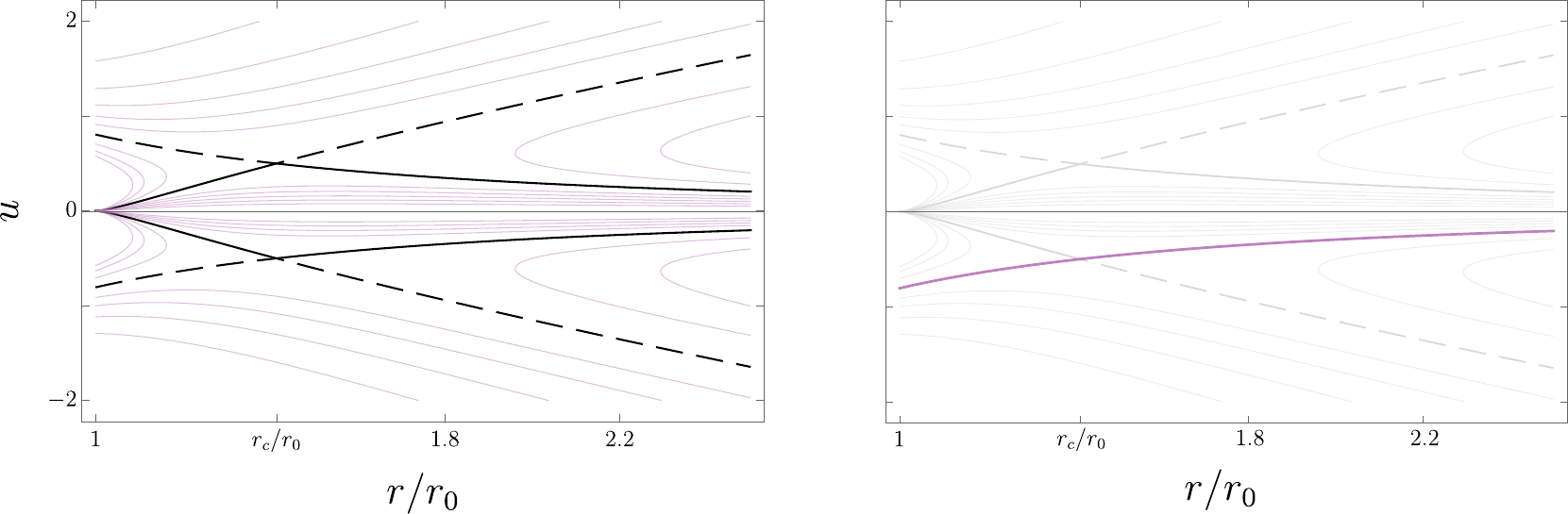}
    \caption{Left: Level curves of the radial component of the four-velocity (purple) as a function of the radial coordinate $r$, expressed in horizon-radius units. The four solution branches are shown in black: $\pm u_1(r)$ (dashed) and $\pm u_2(r)$ (solid), with distinct near-horizon and asymptotic behaviour. Right: Global piecewise solution corresponding to inward accretion matched to a static background.}
    \label{fig:ur}
\end{figure}

For a general radius $r$, eqs.~\eqref{A1} and~\eqref{A2} define a system for the energy density $\rho(r)$ and the radial component of the four-velocity $u(r)$. From this system, one can obtain level curves of $u(r)$ (or $\rho(r)$), shown as contour lines in figure~\ref{fig:ur} for different values of $A_1$ and $A_2$. Physical solutions must pass through the critical point $(r_{\rm cr}, \pm u_{\rm cr})$ (or $(r_{\rm cr}, \rho_{\rm cr})$).

\begin{figure}
    \centering
    \includegraphics[width=0.75\linewidth]{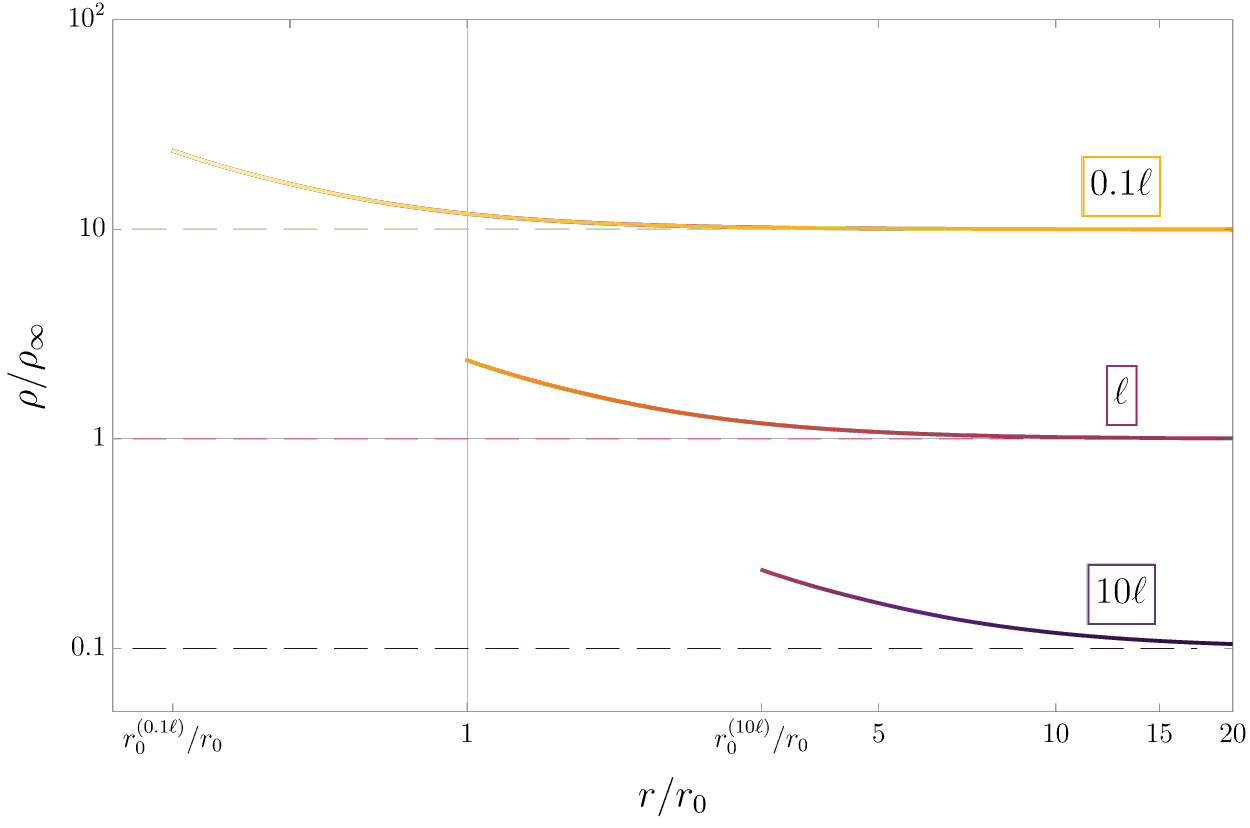}
    \caption{Energy density profiles for fixed mass and cosmic time, shown for different AdS radii. Both axes are normalised to the background energy density $\rho_\infty$ and horizon radius $r_0$ corresponding to the reference scale $\ell$. The lower and upper curves correspond to $\ell_1 = 10\ell$ and $\ell_2 = 0.1\ell$, respectively.}
    \label{fig:comp-dens}
\end{figure}

We find six branches for $u(r)$ that include the critical point, four of which are real in the domain $r \geq r_0$ and are shown in figure~\ref{fig:ur} in black. For accretion, we select the inward-flow solutions with $u < 0$ and analyse the near-horizon and asymptotic behaviour of the two relevant branches. For $r < r_{\rm cr}$, we choose the subcritical branch ($u < u_{\rm cr}$), such that $\lim_{r \to r_0} u(r) < 0$, while for $r \geq r_{\rm cr}$ we select the supercritical branch, satisfying $\lim_{r \to \infty} u(r) = 0$. This choice is shown in Figure~\ref{fig:ur}.
The corresponding energy density profile is
\begin{equation}\label{dens-prof}
    \rho(x)=\frac{16}{9}\left(x^2|u(x)|\right)^{-\frac{4}{3}}\rho_\infty\quad,\qquad x=\frac{r}{r_0}.
\end{equation}
Note that, for a fixed mass, the variable $x$ depends on the choice of the AdS scale, as does $\rho_\infty$, following the scaling
\begin{equation}\label{scaling}
    \rho_{\infty}(\ell_1)=\frac{\ell_2}{\ell_1}\rho_\infty(\ell_2)\quad ,\qquad x(\ell_1)=\sqrt{\frac{\ell_1}{\ell_2}}\,x(\ell_2)\,,
\end{equation}
using eqs.~\eqref{quad-reg} and~\eqref{horizonradius}. In Figure~\ref{fig:comp-dens}, we illustrate the impact of AdS warping on the energy density profile. We consider a black hole of mass $M$ at fixed cosmic time $t < t_c$, characterised by a horizon radius $r_0(\ell)$ and background energy density $\rho_\infty$. We compare two AdS radii, $\ell_1 = 10\ell$ and $\ell_2 = 0.1\ell$, with corresponding quantities $(r_{0,i}, \rho_{\infty,i})$. Using the scaling relations~\eqref{scaling}, we express $(r_{0,i}, \rho_{\infty,i})$ in terms of $(r_0, \rho_\infty)$ and compute the resulting energy density profiles. We find that, for fixed mass, the profile becomes more diluted as the warping of the extra dimension increases.
%
%
%
%
\subsection{Black hole mass evolution}
\begin{figure}
    \centering
    \includegraphics[width=1\linewidth]{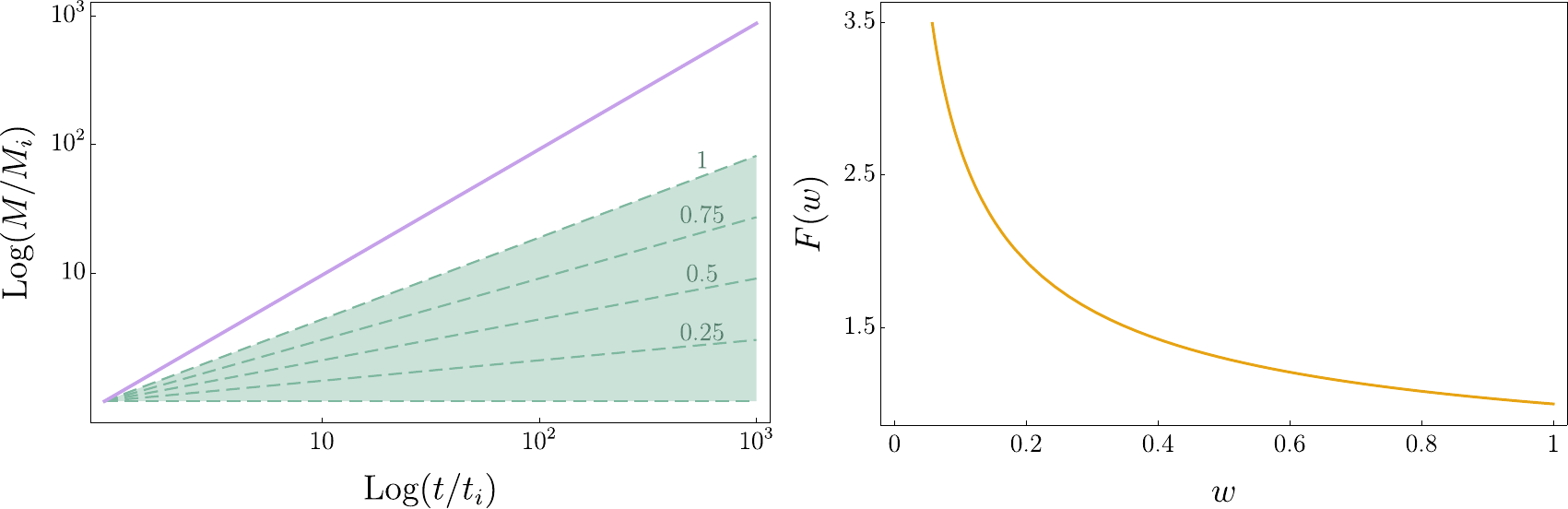}
    \caption{Left: Black hole mass evolution $M(t)$ as a function of cosmic time for $w = 1/3$. The green shaded region shows the range obtained in refs.~\cite{Guedens_2002-2,Majumdar_2003} using Bondi accretion with different efficiencies ($F = 0, 0.25, 0.5, 0.75, 1$), while the purple curve corresponds to our result based on Michel accretion. Both axes are rescaled: the mass is normalised to the initial value $M_i = M(t_i)$, and time is expressed in units of the formation time $t_i$. Right: The factor $F(w)$ as a function of the equation-of-state parameter $w$.}
    \label{fig:comparing-results}
\end{figure}

In order to find the accretion rate, we evaluate eq.~\eqref{A1} at the critical point. Using eqs.~\eqref{eqs-as-w} and~\eqref{rho-c}, we find
\begin{equation}\label{mdot1}
    \dot{M}=4\pi F(w)r_0^2\,\rho_\infty
\end{equation}
with
\begin{equation}
    F(w)=\frac{(1+w)^{\frac{1+w}{2w}}}{2\sqrt{w}}\,,
\end{equation}
which evaluates to $F \simeq 1.54$ for $w = 1/3$. In the regime where Michel’s formalism applies and a barotropic equation of state holds at all radii, $F$ is expected to take this value \emph{independently of the microscopic interactions within the fluid}, as it follows from purely geometric considerations. Interestingly, our result coincides with that of ref.~\cite{Tikhomirov:2005bt}, despite not employing the approximations used there to evaluate $A_1$, and \emph{without} invoking microscopic properties beyond a barotropic radiation equation of state. The behaviour of $F(w)$ is shown in Figure~\ref{fig:comparing-results}.

Note that we recover the same functional form as Clancy et al.~\cite{Guedens_2002-2} and Majumdar~\cite{Majumdar_2003}, although these authors adopt $0 < F \leq 1$ and $F = 1$, respectively. In figure~\ref{fig:comparing-results}, we compare their results (shown in green) with our prediction (purple). The shaded green region represents the range of possible mass evolution functions associated with different values of $F$, interpreted in ref.~\cite{Guedens_2002-2} as an accretion efficiency parameter. In contrast, in our treatment $F$ is fixed by the equation of state of the accreted fluid and decreases as the equation of state softens. We find larger values of $F$  which correspond to significantly enhanced accretion during the quadratic regime.

Using eqs.~\eqref{quad-reg} and~\eqref{horizonradius}, we obtain~\eqref{mdot1} as a function of black hole mass and cosmic time,
\begin{equation}
    \dot{M}=\frac{2F(w)}{\pi}\frac{M}{t}\,.
\end{equation}
Integrating from formation time $t_i$ to an arbitrary time $t < t_c$ yields the black hole mass evolution,
\begin{equation}\label{Mt-result}
    M(t)=M_i\left(\frac{t}{t_i}\right)^{\frac{16}{3\sqrt{3}\pi}}
\end{equation}
for $w=1/3$. 
If we compare eq.~\eqref{Mt-result} with the estimates $M_{\rm prev}(t)$ obtained in refs.~\cite{Guedens_2002-2,Majumdar_2003},
\begin{equation}
    \frac{M}{M_{\rm prev}}=\left(\frac{t}{t_i}\right)^{\frac{2}{\pi}\left(F(w)-F_{\rm prev}\right)}\,,
\end{equation}
we find that that for the maximal accretion rate $F_{\rm prev}=1$, the discrepancy grows with time as $(t/t_i)^{0.34}$ and can reach several orders of magnitude by $t_c$, depending on the formation time. For $F_{\rm prev}<1$, this effect is even more pronounced.

To ensure that this strong growth is consistent with causality, we compare the evolution of the cosmological horizon and the Schwarzschild radius of the accreting black hole. We define a parameter $0 < f_H < 1$ that quantifies their relative size at the formation time,
\begin{equation}\label{fH}
    r_0(M_i)=f_H H^{-1}(t_i)
\end{equation}
and combining eqs.~\eqref{quad-reg},~\eqref{Mt-result} and~\eqref{fH},
\begin{equation}\label{r0-H-1}
    \frac{r_0(t)}{H^{-1}(t)}\simeq f_H\left(\frac{t_i}{t}\right)^{0.51}\,.
\end{equation}
Hence, we find that the horizon radius grows more slowly than the cosmological horizon. We impose only the condition $f_H < 1$, and do not assume $f_H \simeq 1$ as in the conventional scenario, since gravitational collapse in braneworld cosmology requires a more careful treatment (see ref.~\cite{Casadio_2005}).

Equation~\eqref{mdot1} remains valid in the regime $t \gg t_c$, provided the energy density is given by eq.~\eqref{linear-reg}. In this case, integration yields
\begin{equation}
    M(t)=M(t_c)\exp\left[\frac{2F(w)}{\pi}\left(1-\frac{t_c}{t}\right)\right]\,,
\end{equation}
in agreement with ref.~\cite{Guedens_2002-2}. In the regime $t \gg t_c$, the difference between our value of $F$ and the range adopted in the literature is negligible, and the black hole mass remains constant up to a factor of order unity.
\subsection{Strong accretion epoch}
To determine the duration of the strong accretion phase, it is necessary to identify the earliest possible formation time, namely the end of inflation in the RS-II scenario. Establishing this initial epoch enables us to relate final black hole masses to their corresponding initial values and to track their evolution in a quantitatively consistent framework. To this end, we make use of the relation between the tensor ($\mathcal{P}_{\rm \small T}$) and scalar ($\mathcal{P}_{\rm \small S}$) power spectra
\begin{equation}
     \mathcal{P}_{\small \rm T}=r\,\mathcal{P}_{\small \rm S}\,,
\end{equation}
with $r$ the tensor-to-scalar ratio. The tensor power spectrum in braneworld scenarios was first derived in~\cite{Langlois_2000}, with subsequent analyses confirming this result~\cite{Seery_2005,Gorbunov_2001}. In these works it is shown that, relative to the four-dimensional case, the tensor power spectrum acquires an additional factor $\mathcal{F}^2$, such that
 \begin{equation}
      \mathcal{P}_{\rm T}=\frac{2}{\pi^2}\frac{H^2}{M_4^2} \mathcal{F}^2\quad,\qquad \mathcal{F}^2=\left[\sqrt{1+H^2\ell^2}-H^2\ell^2\ln\left(\frac{1}{H\ell}+\sqrt{1+\frac{1}{H^2\ell^2}}\right)\right]^{-1}\,,
 \end{equation}
 resulting in an enhanced gravitational-wave amplitude in the quadratic regime for fixed $H^2$.\,\footnote{Note that, in contrast to~\cite{Langlois_2000}, we set their parameter $\mu$ to $\mu^{-1} = l$; this identification holds only in the limit $\Lambda_4 \simeq 0$.}

\begin{figure}
    \centering
    \includegraphics[width=0.75\linewidth]{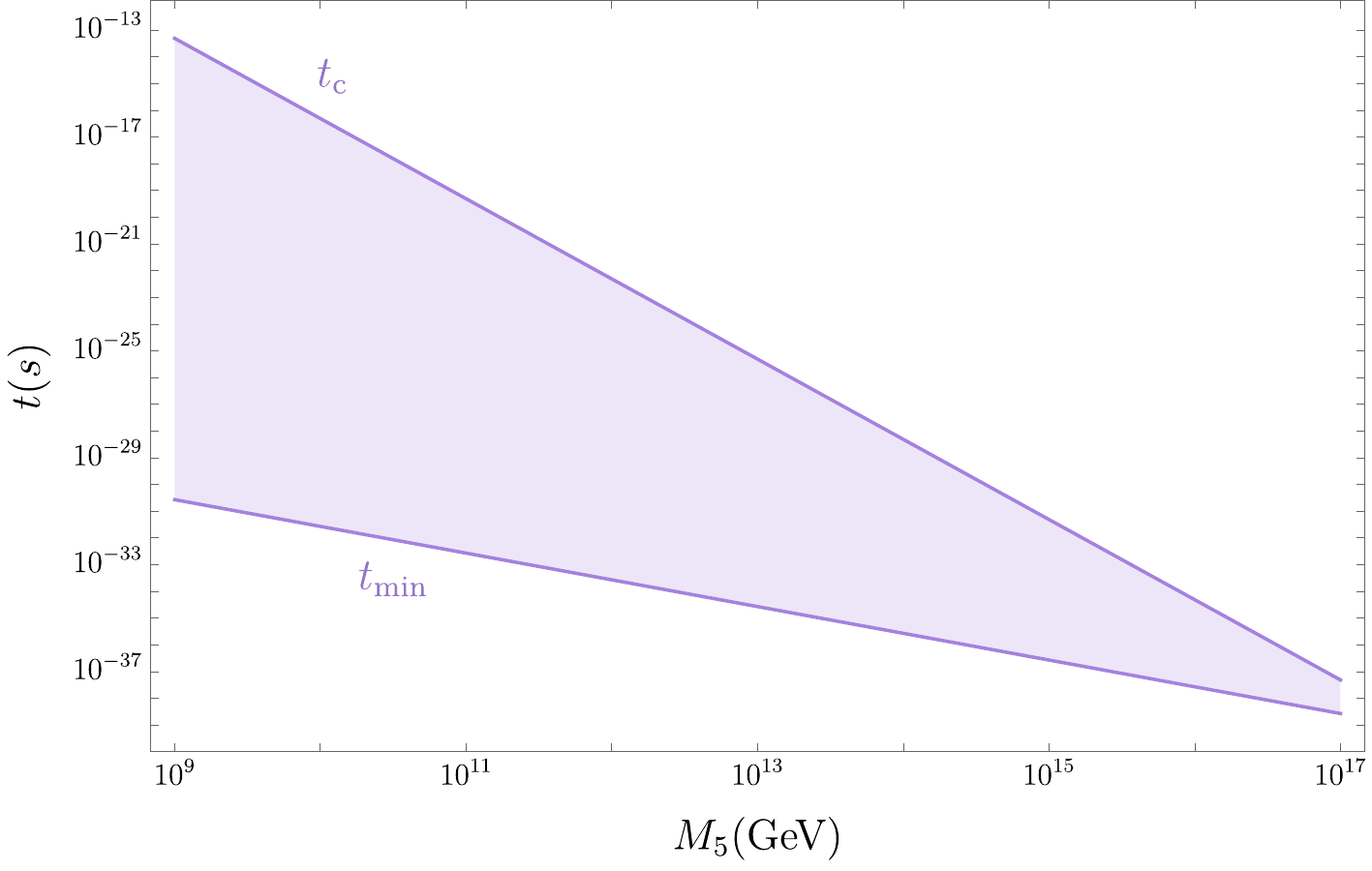}
    \caption{Maximum duration of the strong-accretion epoch for different $M_5$ values. The upper bound is set by the end of the quadratic regime $t= t_c$ and the lower bound is imposed by the scalar power spectrum and tensor-to-scalar ratio observational constraints.}
    \label{fig:duration}
\end{figure}

The function $\mathcal{F}$ simplifies in the quadratic regime,
\begin{equation}
    \mathcal{F}\simeq \sqrt{\frac{3H\ell}{2}}\,,
 \end{equation}
 and $\mathcal{F} \to 1$ for $\rho \ll \lambda$, recovering the four-dimensional result. Observational constraints on $r$ and $\mathcal{P}_{\rm S}$, namely $r < 0.032$ (95\% CL)~\cite{Tristram_2022} and $\mathcal{P}_{\rm S} \simeq 2.1 \times 10^{-9}$~\cite{2020}, then allow us to derive a lower bound on $H^{-1}$. This, in turn, sets a corresponding lower bound on $r_0(f_H)$ at formation for different values of $M_5$. The duration of the strong-accretion epoch for different values of $M_5$ can be found in Figure~\ref{fig:duration}. 

\begin{figure}
    \centering
    \includegraphics[width=0.77\linewidth]{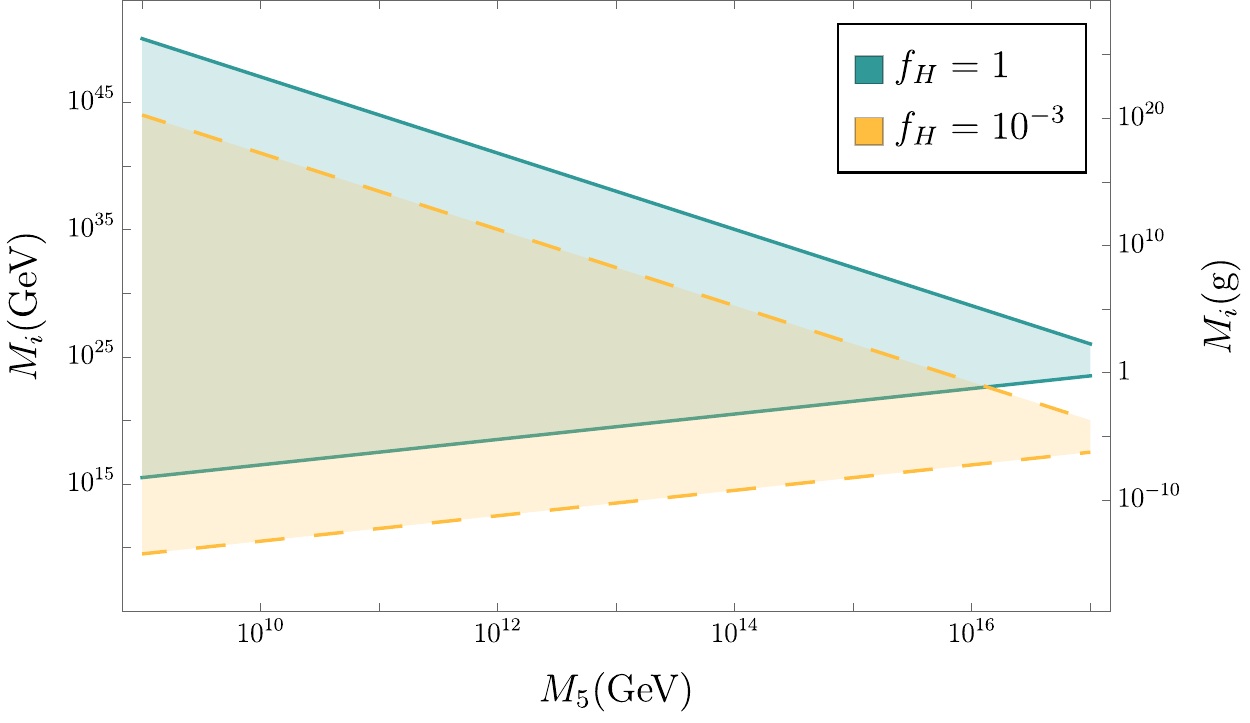}
    \caption{Range of black hole masses at formation as a function of the fundamental Planck scale. The blue shaded region corresponds to $f_H = 1$, while the yellow shaded region shows the case $f_H = 0.1$. In both cases the upper mass value corresponds to $t_i\sim t_c$ and the lower mass value to $t_{i,\rm min}$ imposed by constraints on the power spectrum. The minimum value of the fundamental Planck scale considered is set by current observational constraints.}
    \label{fig:mass-ranges}
\end{figure}

\begin{figure}
    \centering
    \includegraphics[width=0.77\linewidth]{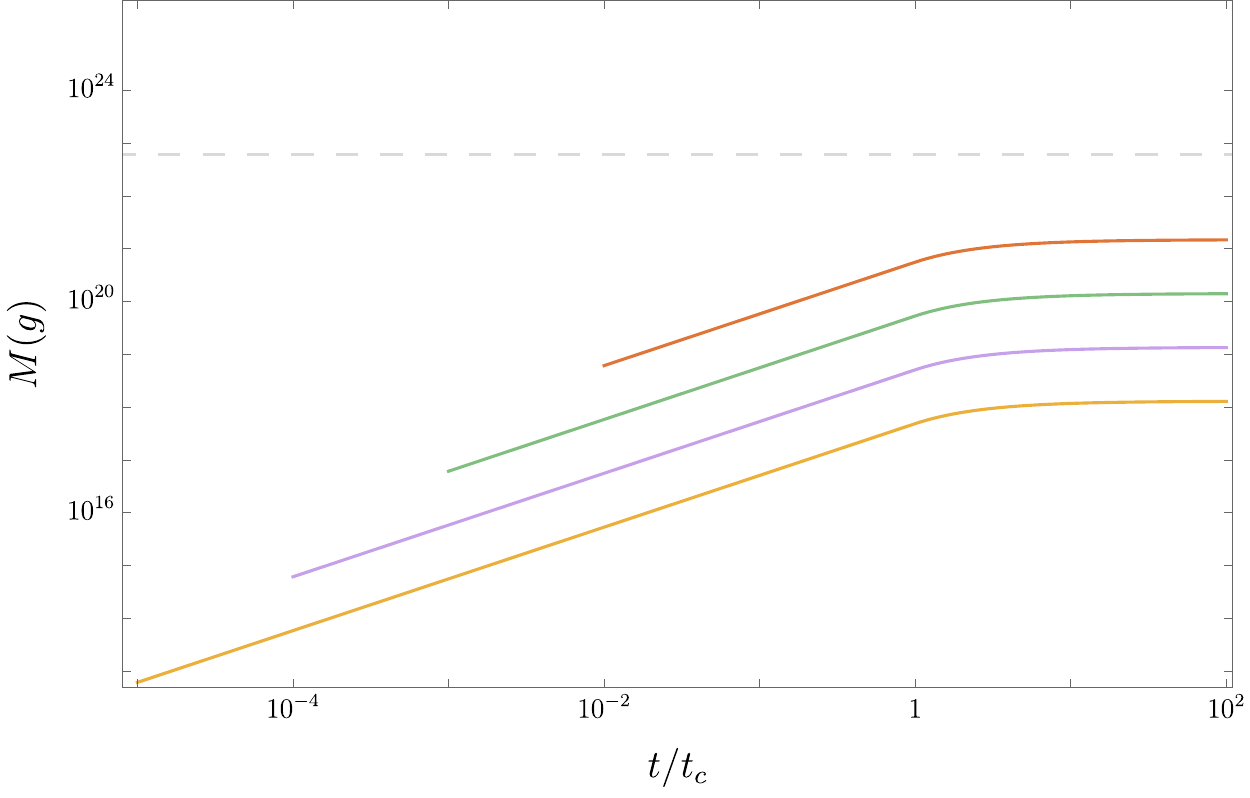}
    \caption{Mass evolution as a function of cosmic time for $(M_5, f_H) = (M_{5,\rm min}, 0.5)$, shown for four different formation times ($t_i=10^{-x}\,t_c$, $x=\{2,3,4,5\}$). Earlier formation leads to a longer accretion phase and correspondingly relative larger final masses. The dashed line indicates the mass for which $r_0 \sim \ell$, marking the breakdown of the Schwarzschild--Tangherlini description.}
    \label{fig:mass-time}
\end{figure}

 In Figure~\ref{fig:mass-ranges}, we present the range of black hole masses if formation takes place during the quadratic regime for two representative values, $f_H = \{1, 0.1\}$. The parameter $f_H$ depends strongly on the formation mechanism (e.g.~in particle collisions, $f_H \ll 1$). In Figure~\ref{fig:mass-time}, we show the mass evolution as a function of cosmic time for $(M_5, f_H) = (10^{6}\,\mathrm{TeV}, 0.5)$, considering four formation times. Earlier formation leads to a longer accretion phase and hence significantly larger masses. To quantify the magnitude of this enhanced accretion, we consider the limiting case of the minimum observationally allowed five-dimensional Planck scale, with primordial black holes forming immediately after inflation: for $f_H \simeq 1$, the initial mass $M_i \sim 10^{-8}\mathrm{g}$ increases to $M(t_c) \sim 10^{8}\mathrm{g}$.

In ref.~\cite{Aldecoa-Tamayo_2026}, we analysed the evolution and observational constraints for a monochromatic primordial black hole mass distribution defined at $t = t_c$. Efficient accretion at $t \ll t_c$ implies that black holes surviving today (e.g.~$M(t_{\rm today},\ell_{\rm max})> 10^{9}\rm g$) may originate from initial masses $M_i < M_4$. As a result, in scenarios such as primordial overdensity collapse, the relation between the primordial power spectrum and present-day black hole masses is significantly modified relative to the standard picture.

\section{Validity of Michel's formalism in braneworlds}\label{section-validity}
Both the Bondi and Michel accretion formalisms rely on strong assumptions: a quasi-static flow embedded in an expanding cosmological background, and an accreting fluid with negligible self-gravity. Assessing the validity of these assumptions is therefore crucial in order to identify the region of parameter space in which the prescription applies. This issue was examined by Ricotti~\cite{Ricotti_2007} for Schwarzschild primordial black holes within the Bondi framework. Here, we extend this analysis to the RS-II scenario and to a fully relativistic treatment.

\subsection{Staticity}
In sections~\ref{section-constants-integration} and~\ref{section-accretion}, we assumed a quasi-static flow, such that the metric, energy densities and four-velocity are time-independent. This approximation is valid when the region near the black hole effectively decouples from the cosmological expansion. We now assess whether this assumption holds in the case under study. To this end, we introduce the parameter
\begin{equation}\label{epsilont}
    \epsilon_t=\frac{t_{\rm acc}}{H^{-1}}
\end{equation}
that allows us to compare the time scales of the system. In~\eqref{epsilont}, $t_{\rm acc}$ is the time scale of accretion, which should be of order
\begin{equation}
    t_{\rm acc}\sim \frac{r_{\rm cr}}{c_{s}}\,,
\end{equation}
since $r_{\rm cr}$ is the scale of the system and $c_s$ is the maximum speed of propagation of information. For this particular case, 
\begin{equation}\label{epsilon-t}
    \epsilon_t=\sqrt{6}\left(\frac{r_0}{H^{-1}}\right)\,.
\end{equation}
Following eq.~\eqref{r0-H-1}, we can conclude that $\epsilon_t<1$ shortly after formation, depending on the parameter $f_H$. 

In figure~\ref{fig:accretion-scale} we compare the radial accretion scales for a fixed mass for different fundamental Planck scales $M_5$ and for the four-dimensional case. 
\begin{figure}
    \centering
    \includegraphics[width=0.65\linewidth]{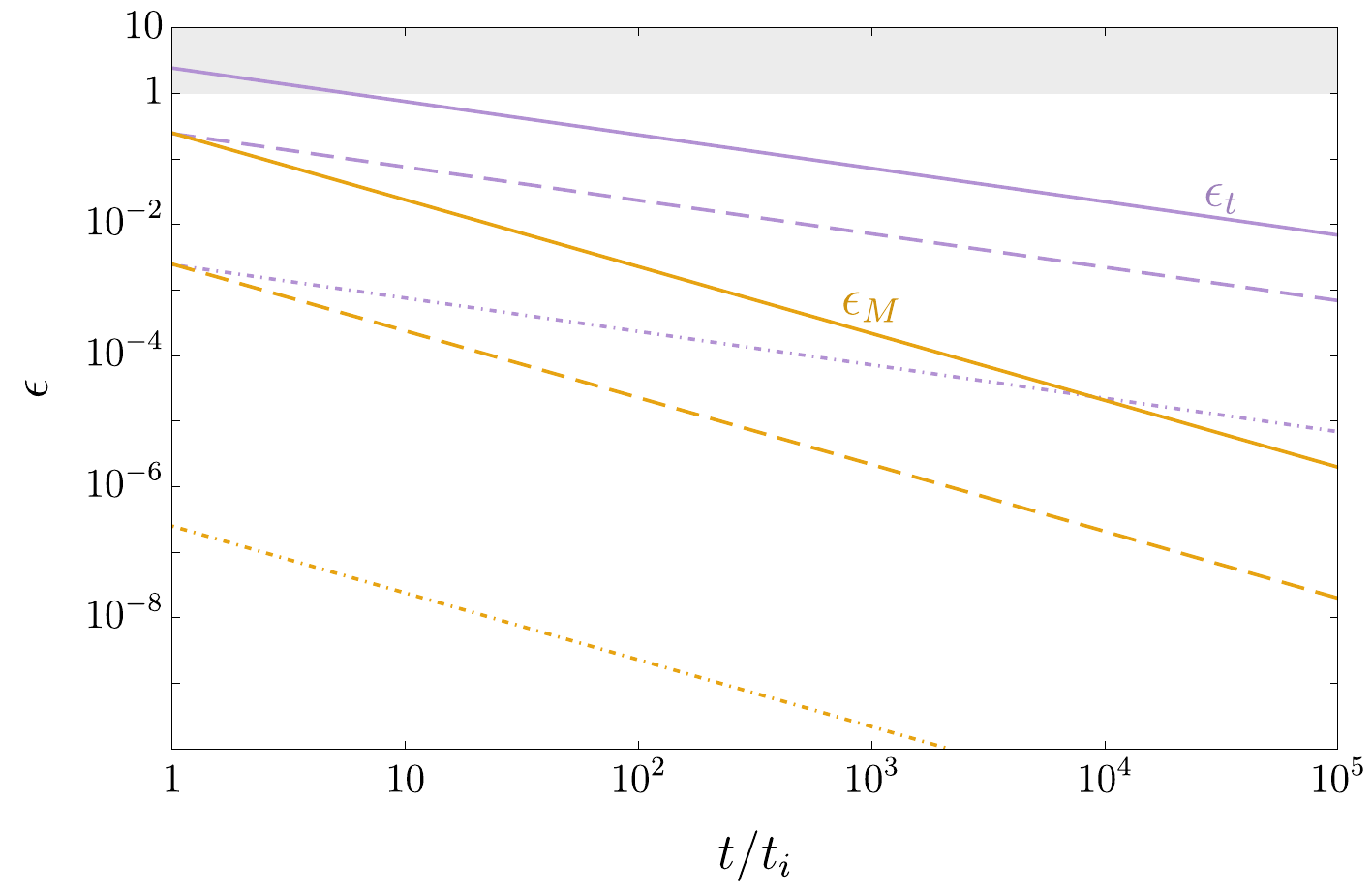}
    \caption{Normalised cosmic time evolution of the staticity and self-gravity parameters, $\epsilon_t$ (in purple) and $\epsilon_M$ (in orange), respectively. for different $f_H$ values ($f_H=1$ solid, $f_H=10^{-1}$ dashed, $f_H=10^{-3}$ dot-dashed). For $\epsilon_t\lesssim 0.1$ and $\epsilon_M\lesssim0.1$, Michel's prescription provides a good approximation.}
    \label{fig:epsilons}
\end{figure}
%
%
\subsection{Test fluid approximation}
In sections~\ref{section-constants-integration} and~\ref{section-accretion}, we assumed that the spacetime metric is given by the Schwarzschild--Tangherlini solution, thereby neglecting the gravitational backreaction of the surrounding fluid.

Although the energy density profile becomes more dilute in the presence of a warped extra dimension, its contribution to the geometry is governed not only by the stress-energy tensor, but also by the additional terms $S_{\mu\nu}$ and $\mathcal{E}_{\mu\nu}$. To assess the validity of this approximation, we introduce an effective mass function—following ref.~\cite{Casadio_2005} in the context of gravitational collapse—to quantify the enclosed gravitational energy ($r_0<r<r_{\rm cr}$),
\begin{equation}\label{eq-meff}
    M_{\rm eff}=4\pi \int_{r_0}^{r_{\rm cr}} \rho_{\rm eff}(r)\,r^2\dd r\,,
\end{equation}
where we have integrated over the spatial volume $\sqrt{h}\,\dd r\dd \theta\dd\phi$, and compare it to that of the black hole's Weyl cloud,
\begin{equation}\label{eq-mbh}
    M_{\rm \mathcal{E}}=4\pi\int_{r_0}^{r_{\rm cr}}\rho_{\mathcal{E}}^{\small \rm BH}\,r^2\dd r\,,
\end{equation}
with $\rho_{\mathcal{E}}^{\small \rm BH}$ provided by~\eqref{rho-E}. If $|M_{\rm eff}|\ll| M_{\mathcal{E}}|$, the perturbation of the metric induced by the fluid can be neglected, and the procedure used above to compute the accretion rate is justified.

In this context, the effective energy density $\rho_{\rm eff}$ must be used in place of the fluid energy density, as the metric receives contributions from additional gravitational sources.

To determine $\rho_{\rm eff}(r)$, we consider a spherically symmetric energy density profile that is approximately decoupled from the cosmic expansion, i.e.\ $\dot{\rho}(r) = 0$, and that asymptotically matches the FLRW background. If the static approximation holds, we may use the non-local integration constant~\eqref{A3} and evaluate it in the asymptotic limit during the quadratic regime. This yields
\begin{equation}\label{rho-E}
\rho_\mathcal{E}(r)=  \frac{3(1+w)}{4\lambda}\left( \rho_\infty-\rho(r)\right)\rho(r)\,.
\end{equation} 
This corresponds to the energy density profile of dark radiation sourced by a spherically symmetric distribution of brane matter in the quadratic regime.\,\footnote{This expression does not account for contributions sourced by bulk matter. Throughout, we assume an empty bulk, so that dark radiation is generated exclusively as a backreaction of gradients in the brane energy density.}

\begin{figure}
    \centering
    \includegraphics[width=0.78\linewidth]{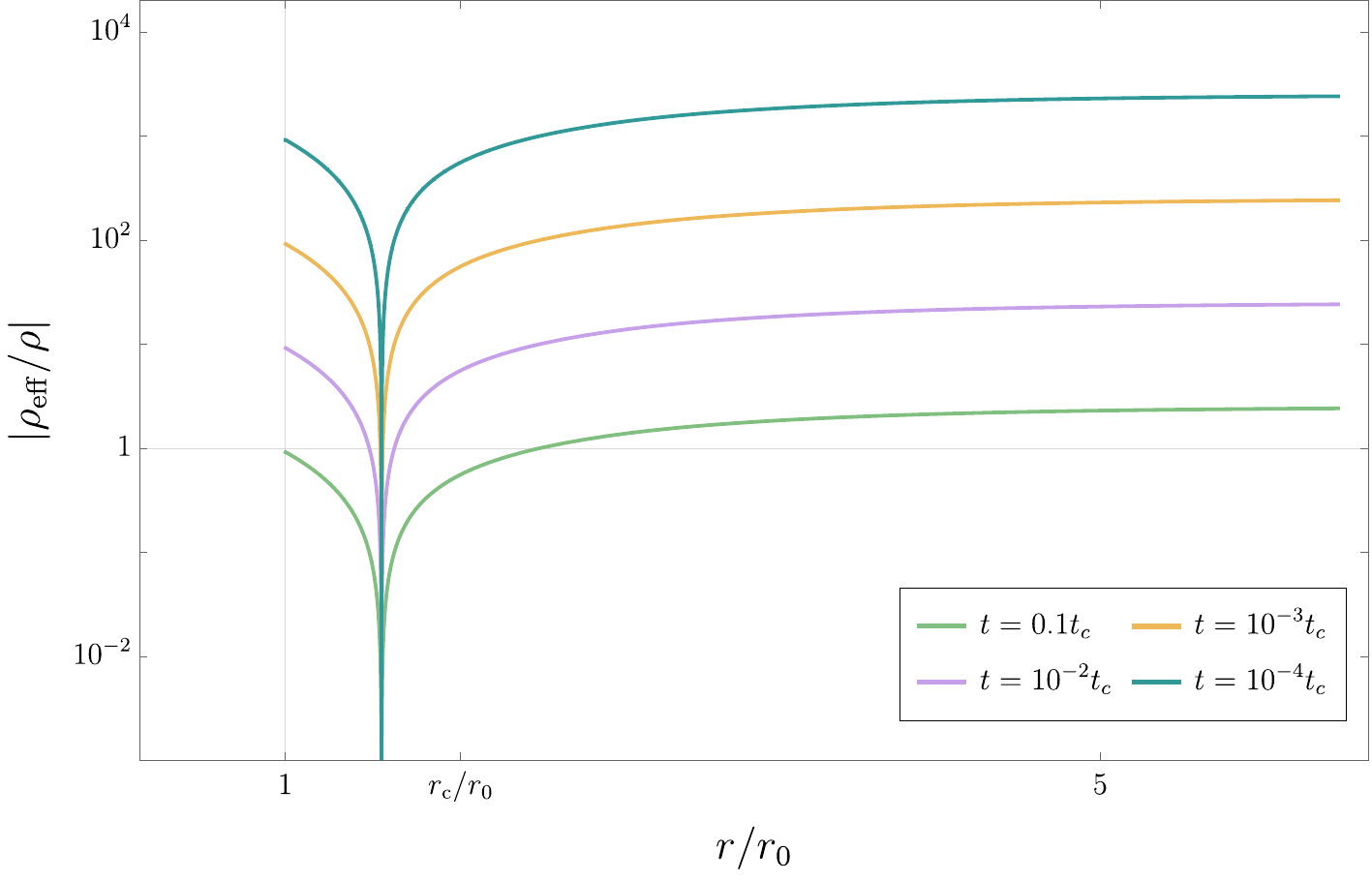}
    \caption{Absolute value of the ratio of the effective energy density and the brane's fluid energy density in the vicinity of a five-dimensional black hole with horizon radius $r_0$ at different cosmic times.}
    \label{fig:ratio-rhos}
\end{figure}

Therefore, the effective energy density~\eqref{rho-eff} in this region takes the form 
\begin{equation}\label{rho-eff-1}
    \rho_{\rm eff}\simeq \frac{\rho(r)}{\lambda}\left(\rho_\infty-\frac{\rho(r)}{2}\right)
\end{equation}
for $p=\rho/3$ and $\rho_\infty\gg\lambda$. In Figure~\ref{fig:ratio-rhos} one can see the absolute ratio $|\rho_{\rm eff}/\rho|$ for different cosmic times. Allow us to highlight two key features:
\begin{itemize}
    \item The ratio depends strongly on cosmic time, with the contributions from $S_{\mu\nu}$ and $\mathcal{E}_{\mu\nu}$ becoming significant at early times.
    \item The quadratic correction dominates in the asymptotic limit $r \to \infty$. As the density profile steepens, the negative contribution from $\rho_{\mathcal{E}}$ becomes dominant, leading to $\rho_{\rm eff} < 0$, with large negative values $|\rho_{\rm eff}|/\rho \sim t_c/t$.
\end{itemize}
Evaluating the integrals~\eqref{eq-meff} and~\eqref{eq-mbh}, we obtain an estimate of the ratio
\begin{equation}\label{epsilon-M}
   \epsilon_M= \Bigg|\frac{M_{\rm eff}}{M_{\mathcal{E}}}\Bigg|\simeq 0.2\,\left(\frac{r_0}{H^{-1}}\right)^2\,.
\end{equation}
Together with eq.~\eqref{r0-H-1}, this justifies neglecting the backreaction of the surrounding fluid at all cosmic times. The evolution of $\epsilon_M$ and $\epsilon_t$ is shown in figure~\ref{fig:epsilons}. As shown in this figure, for $f_H \simeq 1$, the parameter $\epsilon_t$ exceeds unity at early times. This suggests that, in scenarios where black holes form with event radii comparable to the cosmological horizon, the Michel approximation may not accurately describe the accretion process immediately following formation.
%
%
\section{Conclusions}\label{section-conclusions}
In this work, we have presented a comprehensive analysis of accretion onto five-dimensional Schwarzschild black holes in the RS-II framework, adopting the fully relativistic treatment introduced by Michel~\cite{Michel1972}. We ensured consistency with the higher-dimensional setting at each step and quantified the impact of braneworld corrections.

By performing a complete analysis of scalar perturbations, we determined the propagation speed of pressure-supported disturbances in the accreting fluid, which is essential for assessing the sonic properties and relevant time scales of the system. We find that this speed can be approximated by the thermodynamic sound speed $c_s$ in the quadratic regime, but only locally ($k\eta \to 0$) and for a barotropic equation of state.

We then computed the critical radius of Schwarzschild--Tangherlini black holes, obtaining $r_{\rm cr} = \sqrt{2}\,r_0$ for $w = 1/3$, in agreement with ref.~\cite{Tikhomirov:2005bt}, although derived using a different conservation equation that treats the fluid as highly relativistic. This allowed us to determine the velocity and energy density profiles, revealing that, unlike in the standard four-dimensional case, the fluid remains subsonic at the critical point.

Using these results, we derived the black hole mass evolution. While its functional dependence matches that found in previous studies~\cite{Guedens_2002-2,Majumdar_2003,Tikhomirov:2005bt}, we obtain a different value and interpretation for the parameter $F$. In particular, we find significantly stronger accretion than previously reported, while maintaining consistency with causality.

By imposing observational constraints on the tensor power spectrum, we established a lower bound on the end of inflation in the RS-II scenario, thereby fixing the earliest possible time for primordial black hole formation in this particular framework. Combining this with the mass evolution, we find that earlier formation leads to a prolonged strong-accretion phase and enhanced growth. Strong AdS warping implies an earlier end of inflation and a delayed transition out of the quadratic regime, thereby extending the epoch of strong accretion. In contrast, weaker warping results in a later end of inflation and an earlier transition, reducing the duration of the accretion phase.

We show that, if formed sufficiently early, black holes with initial masses below the effective four-dimensional Planck scale may survive until today, evolving as discussed in our previous work~\cite{Aldecoa-Tamayo_2026}, and thus constitute viable dark matter candidates.

Finally, we assessed the validity of the assumptions underlying our accretion framework. Despite the enhanced radial accretion scale relative to the four-dimensional case and thus the longer accretion time scales, the quasi-static approximation is still robust in braneworlds. In contrast, in $(3+1)$ dimensions this approximation is only valid in a restricted parameter space. Moreover, the test-fluid approximation remains well satisfied throughout the quadratic regime, as the energy density of the matter fields decreases in the RS-II scenario and the corresponding contribution to the enclosed gravitational energy remains subdominant compared to that of the Weyl cloud surrounding the black hole.

We conclude that strong accretion significantly alters the mapping between the conditions required for primordial black hole formation and the resulting mass function at later times, compared to the standard scenario. In particular, cosmological conditions associated with specific formation mechanisms---such as peaks in the primordial power spectrum in overdensity collapse---correspond to markedly different scales for a given present-day mass. As discussed in our previous work~\cite{Aldecoa-Tamayo_2026}, the present-day properties of such black holes, including their temperature and evaporation rate, are likewise substantially modified.

We emphasise that the formation time and the size of the black hole relative to the cosmological horizon depend sensitively on the formation mechanism and the details of the collapse. A more precise characterisation therefore requires a dedicated treatment of these processes and their inclusion in the calculation.

\acknowledgments
IAT thanks R.~Emparan, J.~Garriga and C.~Germani for insightful discussions on the interplay between the five- and four-dimensional descriptions, D.~Blas for useful references on information propagation, K.~Koyama for exchanges on the wave equation calculations, and C.~Requilé for assistance with some of the mathematical tools employed in this work. IAT also thanks the Institute of Cosmos Sciences of the University of Barcelona for its hospitality during part of this project. The research of IAT is supported by an STFC studentship. CB and DS are supported by the STFC grants ST/X001040/1 and ST/X000796/1.


\bibliographystyle{JHEP}
\bibliography{biblio.bib}

\end{document}